\documentclass[journal]{IEEEtran}
\usepackage{amssymb}
\usepackage{graphicx}
\usepackage[cmex10]{amsmath}
\usepackage{subfigure}
\usepackage{times}
\usepackage{comment}
\usepackage{multirow}
\usepackage{amsfonts}
\usepackage{txfonts}
\usepackage{cite}
\usepackage{listings}
\usepackage{xcolor}
\usepackage{url}

\begin{document}
%
\title{Perceptual Quality of Video with Periodic Frame Rate and Quantization Variation--Subjective Studies and Analytical Modeling}

%
%


\author{Yen-Fu Ou,
	    Wenzhi Lin,
	    Huiqi Zeng,
	    Yao Wang\thanks{Y.-F. Ou, W. Lin, H. Zeng, and Y. Wang are with the Polytechnic Institute of New York
University, Brooklyn, NY 11201 USA (e-mail: you01@students.poly.edu;
mosquitolwz@gmail.com; yoyoapply@gmail.com; yao@poly.edu).}
}

\maketitle
%

\begin{abstract}
In networked video applications, the frame rate (FR) and quantization stepsize (QS) of a compressed video are often adapted in response to the changes of the available bandwidth. It is important to understand how do the variation of FR and QS and their variation pattern affect the video quality. In this paper, we investigate the impact of temporal variation of FR and QS on the perceptual video quality. Among all possible variation patterns, we focus on videos in which two FR's (or QS's) alternate over a fixed interval. We explore the human responses to such variation by conducting subjective evaluation of test videos with different variation magnitudes and frequencies. We further analyze statistical significance of the impact of variation magnitude, variation frequency, video content, and their interactions. By analyzing the subjective ratings, we propose two models for predicting the quality of video with alternating FR and QS, respectively, The proposed models have simple mathematical forms with a few content-dependent parameters. The models fit the measured data very well using parameters determined by least square fitting with the measured data.
We further propose some guidelines for adaptation of FR and QS based on trends observed from subjective test results.
\end{abstract}

\section{Introduction}\label{sec:Intro}


In wireless video streaming, due to the limited sustainable bandwidth of a receiver, a video often has to be coded (or transcoded or extracted from a scalable stream) at reduced frame rate (FR), reduced frame size (FS), and/or increased quantization stepsize (QS), so that each coded frame has adequate quality. A critical issue is how to choose the appropriate spatial, temporal and amplitude resolutions (STAR), so as to achieve the best trade-off between picture quality and motion fluidity in the delivered video (Note that amplitude resolution is inversely related to QS). Another challenging problem is that the sustainable bandwidth of a network link often fluctuates in time, calling for adaptation of STAR. One naive approach would be to find the STAR that optimizes the perceptual quality over each short time duration based on the available instantaneous bandwidth. This may however create a video with rapidly fluctuating STAR, which may be annoying to the viewer. For example, variation in frame rate can cause visually annoying jitter artifacts.  It is important to understand how does the variation of the  STAR, individually and collectively, affect the perceived quality. Such understanding would enable us to impose proper constraints on the variation of the STAR, when adapting the STAR based on the time-varying  bandwidth.

Take for example a hypothetical case where the available bandwidth alternates between $R_l$ and $R_h$, and the frame rates that can lead to the best perceived quality for constant rate video at $R_l$ and $R_h$ are $t_l$ and $t_h$, respectively. In this situation, is it better to  code the video with alternating FR's of $t_l$ and $t_h$, or would it be better to stay at $t_l$? More generally, one may want to vary not only the FR, but also the FS and QS to meet the instantaneous rate constraints.

There have been several studies regarding the influence of temporal and amplitude resolutions, individually or jointly, on the perceptual quality \cite{G1070, QualityMetricTQuantBitRate, yenfu_csvt, TemporalMobileVideoBroadcasting, PerceptualTemporalQualityMetrics, Baroncini_2006, Moorty_Mobile_VQA_2012,yenfu_icassp, yenfu_QPVar_icip}. Some of these works (e.g.~\cite{G1070, QualityMetricTQuantBitRate, yenfu_csvt}) consider the cases where the FR and QS are fixed in the entire video, whereas some  (e.g.~\cite{TemporalMobileVideoBroadcasting, PerceptualTemporalQualityMetrics}) consider the impact of FR variation, due to non-uniform and bursty packet losses. Authors in both ~\cite{TemporalMobileVideoBroadcasting, PerceptualTemporalQualityMetrics} proposed quality models based on their subjective quality assessment, however, the model in~\cite{TemporalMobileVideoBroadcasting} involves too many parameters (a total of 4 fixed constants) and the other~\cite{PerceptualTemporalQualityMetrics} predict the quality index by non-linearly combining several sub functions, which can be used to measure different temporal features of video contents. Authors in~\cite{Baroncini_2006} proposed a variable frame rate
control scheme, which adapts the frame rate under fluctuating bandwidth environment. Although this scheme is developed by observing the human preference between fixed and variable frame rate videos they did not provide any mathematical function forms regarding the quality metrics. They provided guidelines for video adaptation instead. In~\cite{Moorty_Mobile_VQA_2012}, authors reported subjective test results for test videos with periodic variations of QS, while the FR and FS are fixed. However, they only considered QS variation under a fixed, relatively slow variation frequency (change every 5 sec), and no analytical models are proposed.

We have conducted subjective tests evaluating the impact of FR variations when QS and spatial resolution are fixed.
Among the many possible patterns of temporal variations, we consider the simple case where the FR alternates between $t_l$ and $t_h$, with each FR staying over a constant time duration Fz. See Fig.~\ref{fig:QPFRVar_test} for an illustration of how the FR and the corresponding video rate changes in time in our test sequences. We study the effect of $t_h$, $t_l$, their difference (variation magnitude) as well as the effect of Fz (inversely related to variation frequency) on the perceived quality. This study directly addresses the questions we raised for the hypothetical example given earlier. But it can also shed lights for more complicated cases where the FR may vary among more than two levels and the variation may not follow a periodic pattern. We also conducted a parallel study of videos with periodic QS variation  between $q_l$ and $q_h$, and investigated the impact of $q_l$, $q_l$, and Fz on the perceived quality. Preliminary results of these studies have been reported in~\cite{yenfu_icassp, yenfu_QPVar_icip}.

In this paper, we report our subjective test results as well as analytical quality models for videos with FR and QS variations, respectively. Section~\ref{sec:SQmeasurement} describes our subjective test configurations. Section~\ref{sec:Test_results_FRVar} presents our study for videos with periodic variation of FR, including subjective test results, and proposed model that predicts the perceived quality based on $t_l$, $t_h$ and Fz. Section~\ref{sec:Test_results_QPVar} presents the similar set of results, but for videos with QS variation. Combining the two subjective test results in aforementioned two sections, Sec.~\ref{sec:FRQPVar_cmp} compares quality of videos with FR and QS variations, respectively, under the same bit rate variation.
Section~\ref{sec:ANOVA_test} investigates the statistical significance of impact of FR/QS variation and video content on perceptual quality using the ANOVA technique. Finally Section~\ref{sec:Conclusion} summarizes our main findings and propose several guidelines for video adaptation based on our findings.


\vspace{-.in}
\section{Subjective Test Setup}\label{sec:SQmeasurement}
\subsection{Testing Material}

Our experiment is conducted using five video source sequences,
Akiyo, Foreman, Football, Ice, Waterfall, all in CIF $(352\times 288)$
resolution and at frame rate 30 fps originally 10 seconds long, which are chosen from
JVT (Joint Video Team) test sequence pool~\cite{JVT_site}.
All these sequences are coded using JVT scalable video model
(JSVM912)~\cite{JSVM}. For each sequence, one bitstream is generated with five temporal layers, with corresponding FR of 1.875, 3.75, 7.5, 15, and 30Hz , and each
temporal layer in turn has five quality layers created with QP equal
to 28, 32, 36, 40 and 44 (with corresponding to QS = 16, 25, 40, 64, 102), respectively, using the coarse grain scalability (CGS) without QS cascading. The GOP size is set to 16 with only the first frame coded as in the I mode. Hierarchical-B structure is used to provide temporal scalability. For motion estimation, the FastSearch mode is enabled with maximum search range of 16 for full-pel search. We use SAD (Sum of Absolute Difference) as cost function for both full-pel and sub-pel. The entropy coding method is CAVLC. The other encoding configurations follow the default settings in JSVM.
\begin{table}[t]
\centering
\caption{Testing configuration for frame rate variation}
\label{tab:testing_points_FRVar}
\begin{tabular}{c|c|c|c}
\hline
QS & Fz&  $t_h$(Hz) & $t_l$(Hz)\\
\hline
\hline
\multirow{4}{*}{16} & \multirow{4}{*}{1/2/3 sec} & 30 & 30/15/7.5 \\
~ & ~ & 15 & 15/7.5 \\
~ & ~ & 7.5 & 7.5 \\
\hline
\end{tabular}
\end{table}
%

%
\begin{table}[htp]
\centering
\caption{Testing configuration for QS variation}
\label{tab:testing_points_QPVar}
\begin{tabular}{c|c|c|c}
\hline
FR & Fz&  ${\rm QS}_b$ & ${\rm QS}_v$\\
\hline
\hline
\multirow{3}{*}{30} & \multirow{2}{*}{1/2/3 sec} & 16 & 16/25/40/64/102 \\
~ & ~ & 40 & 25/40/64/102 \\
\cline{2-4}
~ & 3 sec & 102 & 25/64/102 \\
\hline
\multicolumn{4}{p{6 cm}}{*Note that QS values of 16, 25, 40, 64, and 102, correspond to QP levels of 28, 32, 36, 40, and 44, respectively.}\\
\end{tabular}
\end{table}

%

\begin{figure}[t]
\vspace{-.in}
\centering
  \includegraphics[scale=0.3]{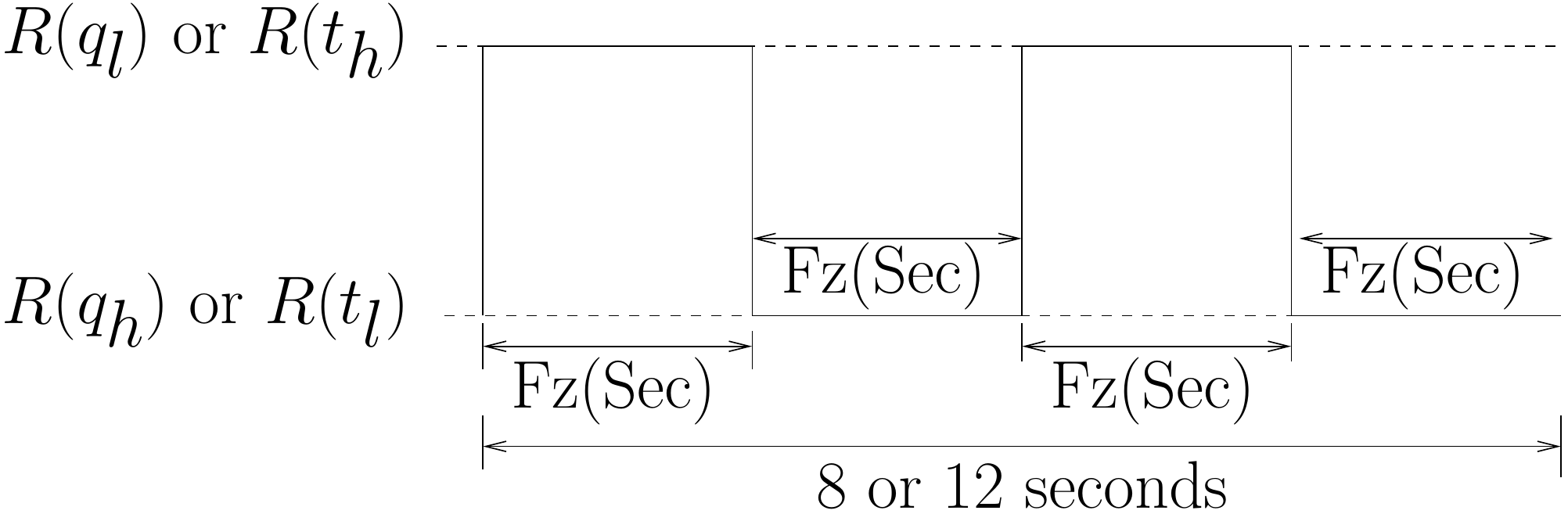}
\caption{The Variations of QS or FR for a video}\label{fig:QPFRVar_test}
\vspace{-.in}
\end{figure}

Two different experiments, examining quality impact of FR and QS variation, were conducted. For temporal variation, as shown in Fig.~\ref{fig:QPFRVar_test}, we generate videos in which two frame rates
switch back and forth periodically through the entire video with changing interval (Fz) of 1, 2, and 3 seconds. The QS is fixed at 16. Let $t_h$ and $t_l$ denote the higher and lower FR of the video.
Table \ref{tab:testing_points_FRVar} details all the test configurations, which leads to a total of 90 processed (encoded and decoded) video sequences (PVS).
For QS variation, we fix FR to 30Hz but allow QS to switch back and forth periodically through the entire video with Fz of 1, 2, and 3 seconds. ${\rm QS}_b$ and ${\rm QS}_v$ denote the base (beginning) QS and deviated QS and the combinations of ${\rm QS}_b$ and ${\rm QS}_v$ are summarized in Tab.~\ref{tab:testing_points_QPVar}. As a result, there are a total of 130 PVS's.

We choose to examine the variation intervals of 1s, 2s, and 3s only, because this range represents most interesting cases to be studied, at least for an initial investigation. It is worth noting that, in video streaming, through the use of buffers, video rate variation can typically be limited to be no faster than changing every second. It is our hope that this initial study can provide important insight regarding the impact of the variation frequency and frequency magnitude on the perceptual quality.

\subsection{Subjective Test Configuration}
The subjective quality assessment is carried out by using a protocol similar to
ACR (Absolute Category Rating) described in
\cite{RecP910}. Basically, each viewer is presented a series of video in a random order, and the viewer is asked to give overall rating of each video in the range of 0 to 100. Each test for one subject consists of two sessions, a training session and a test session. The training session (about 2 minutes) is used for the subject to accustom him/herself to the rating procedure and ask questions if any. The PVS's in the test session (about 12 minutes) are ordered randomly so that each subject sees the video clips in a different order. Each test session contains only a subset of PVS's with either FR or QS variations only. Most of the viewers are engineering students from Polytechnic Institute of New York University, with age 21 to 33. There are on average 22 ratings for each PVS with QS variations, and 20 ratings for each PVS with FR variations. Details regarding each experiment are can be found in~\cite{yenfu_icassp, yenfu_QPVar_icip}.


\subsection{Data Post Processing}
The raw ratings are converted to Z-scores~\cite{AM_zscore} based on the mean and standard deviation of all the scores of each viewer.
In order to remove ``noisy'' ratings or outliers, we adopted, two post screening methods in concatenation. We first perform BT.500-11 post screening method~\cite{BT500} in Z-score domain to remove all ratings by certain viewers because their ratings are outside the range of the majority of the viewers. On average, one viewer is eliminated for each PVS. We then conduct the second step to the remaining ratings in the raw score domain using a ratio/averaging method. Basically we make use of the fact that a video coded at a lower FR (or higher QS) would not have a rating higher than a video coded at a higher FR (or lower QS), if the viewer's judgement is consistent. Therefore, we remove all the ratings by a viewer for the same source video if this viewer violated this expected consistency test for more than a certain number of PVS's for this source. More details can be found in~\cite{yenfu_icassp, yenfu_QPVar_icip}. After this step, on average, 17 ratings remain for each PVS.

\begin{figure*}[ht]
\centering
  \includegraphics[scale=0.4]{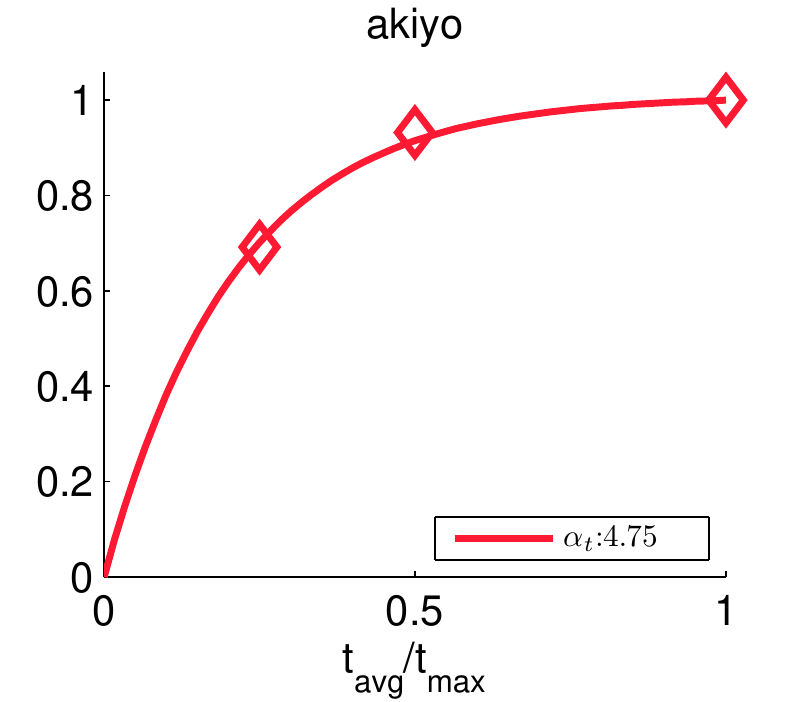}
  \includegraphics[scale=0.4]{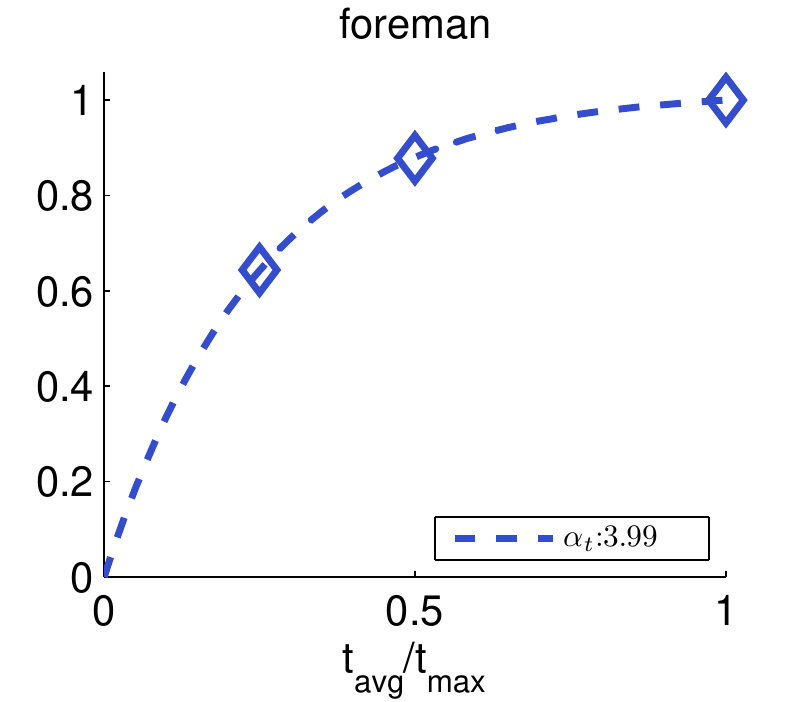}
  \includegraphics[scale=0.4]{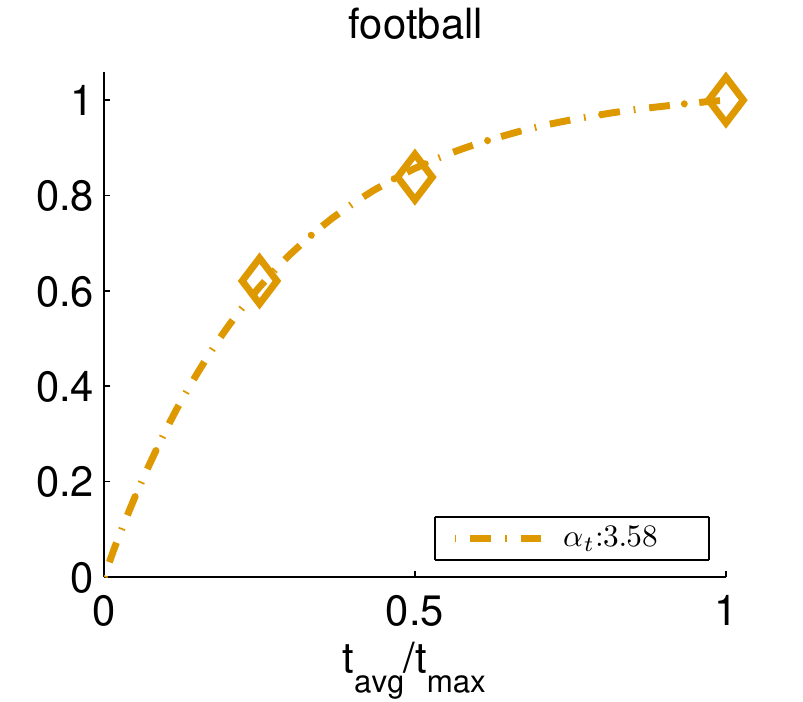}
  \includegraphics[scale=0.4]{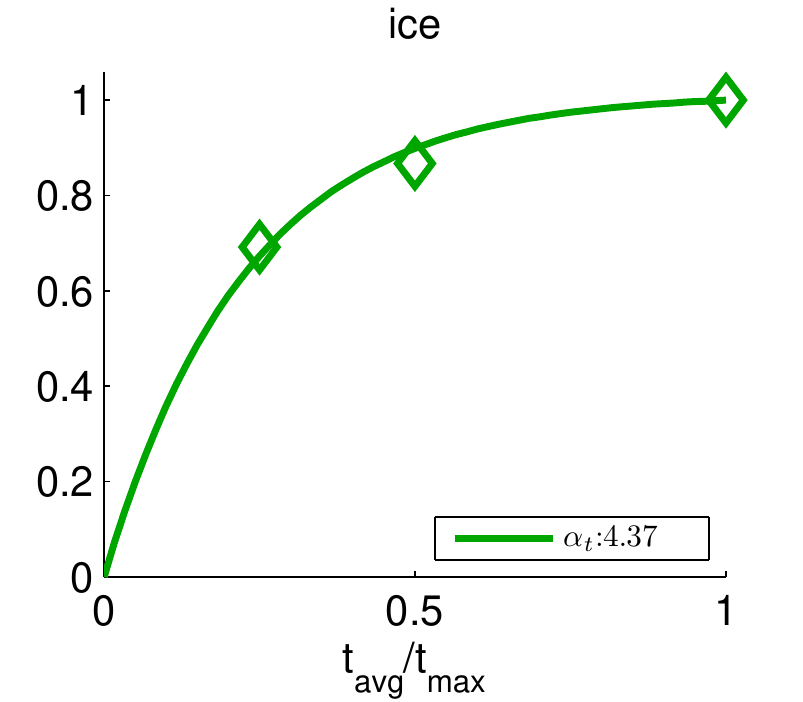}
  \includegraphics[scale=0.4]{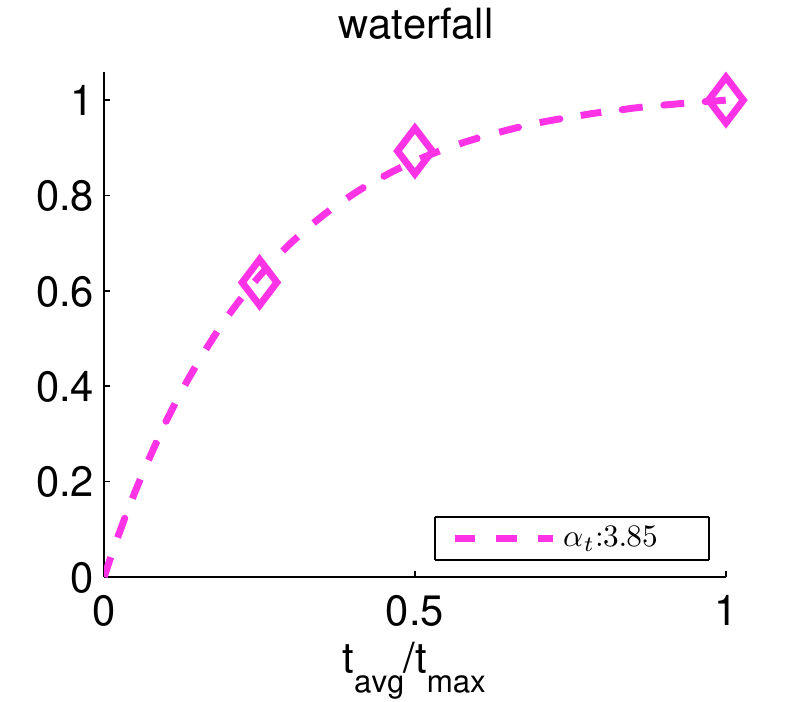}
 \caption{Constant FR case: Q($t,t$) vs. normalized FR ($t/t_{\max}$). Points are the measured data and curves are obtained using Eq.~(\ref{eq:FRVar_MNQT_model}) with PCC=$0.995$, RMSE=$0.013$. Parameter $\alpha_t$ for each sequence is obtained by least squares fitting.}
 \label{fig:FRVar_pMOS_NQT_avgFR}
\end{figure*}


\begin{figure*}[ht]
\centering
  \includegraphics[scale=0.4]{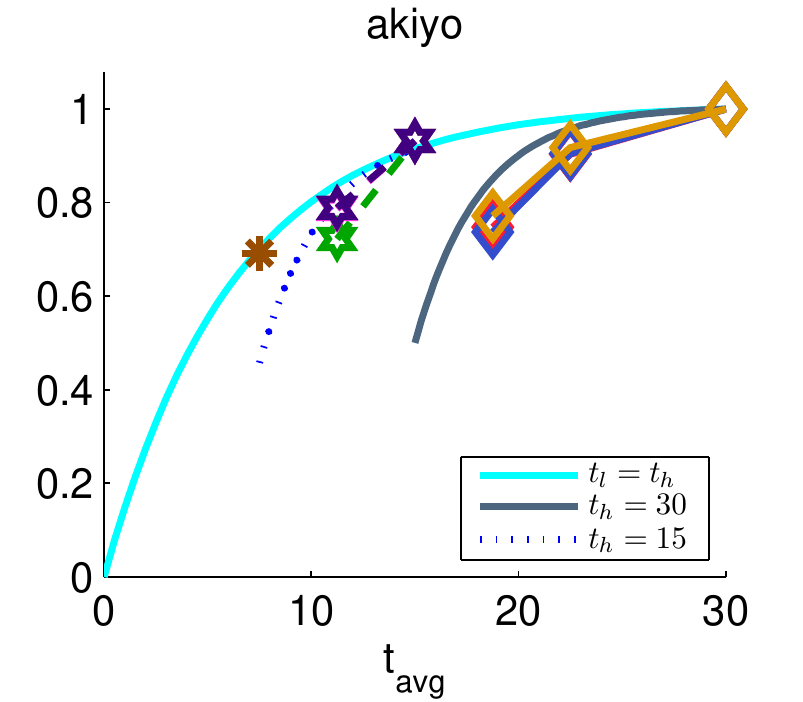}
  \includegraphics[scale=0.4]{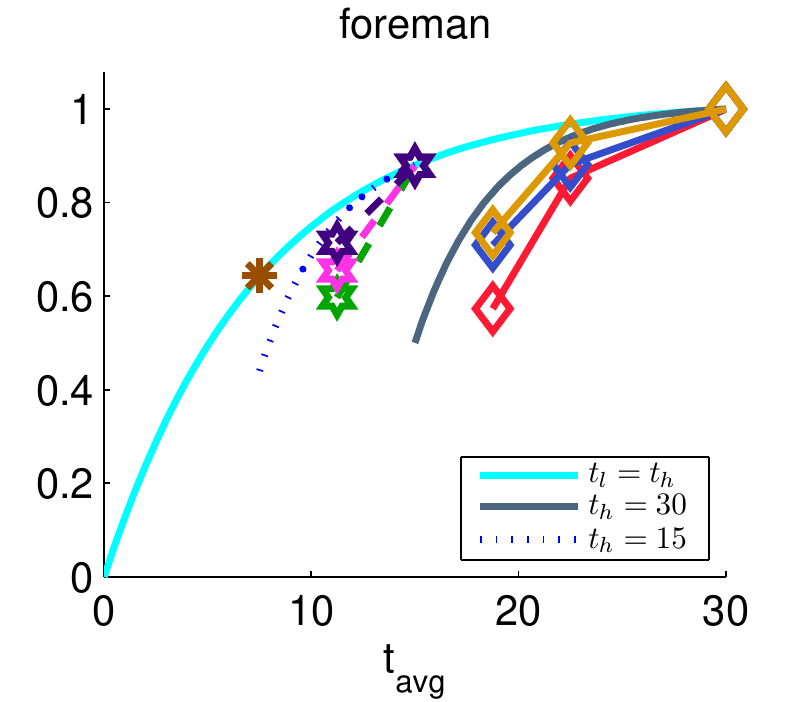}
  \includegraphics[scale=0.4]{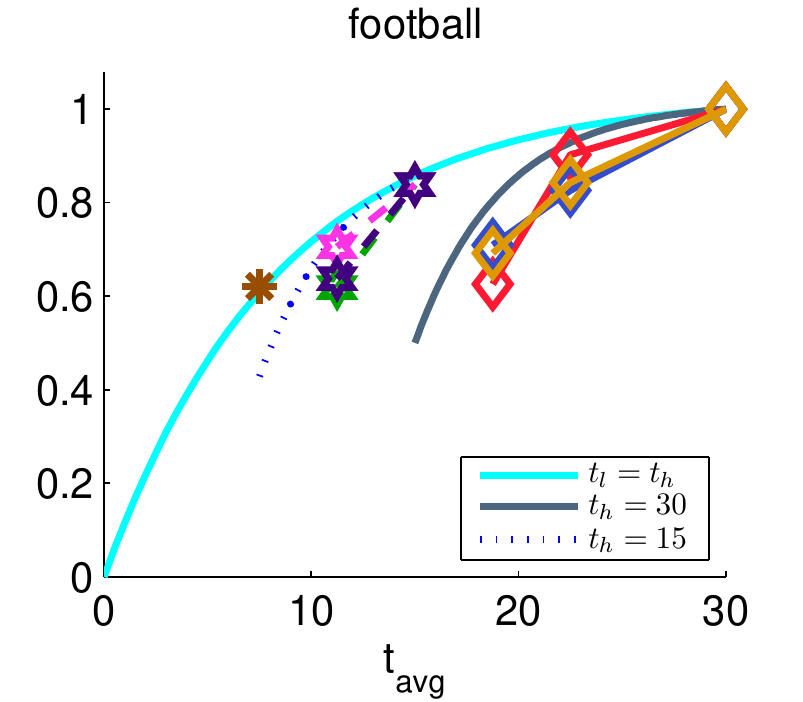}
  \includegraphics[scale=0.4]{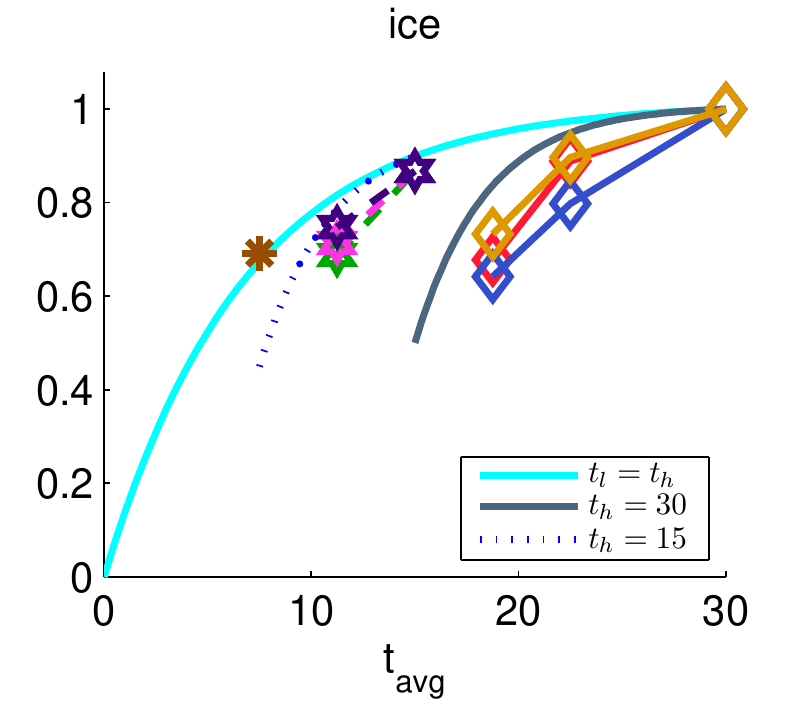}
  \includegraphics[scale=0.4]{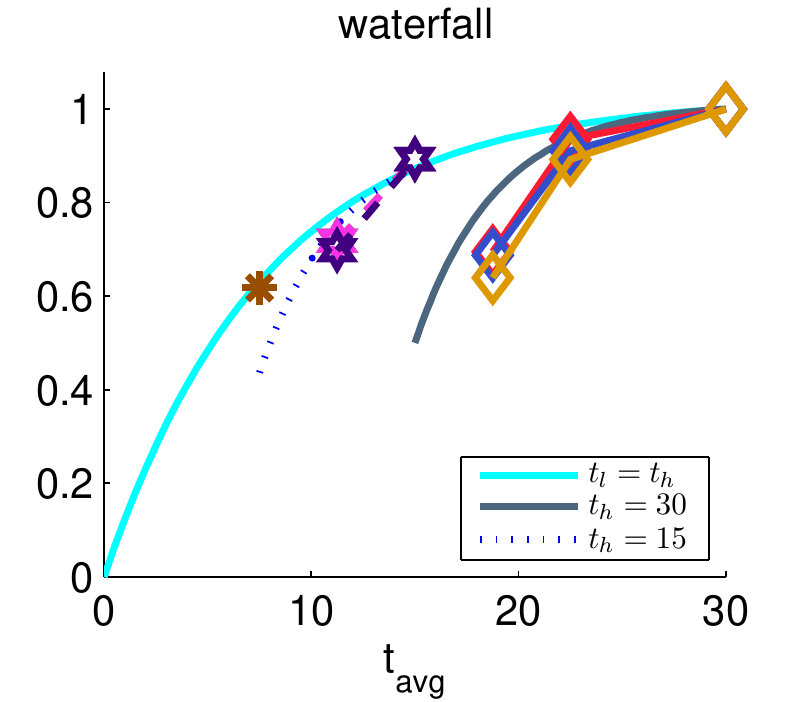}
  \includegraphics[scale=0.35]{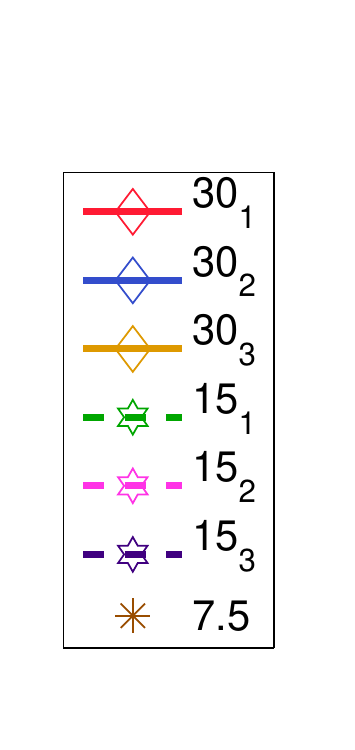}
 \caption{Q($t_h,t_l$) vs. average FR $t_{avg}$. Points are the measured data with different markers and colors corresponding to different $t_h$  and Fz, as indicated in the legend to the right. For example, the points connected by the red line indicated by $30_1$ indicate data points corresponding to different $t_l$, with $t_h$=30, Fz=1 sec. The cyan curves are predicted quality by (\ref{eq:FRVar_MNQT_model}) for constant FR's. The navy solid and blue dotted lines are obtained from (${\rm MNQT}_c$($t_h$)+${\rm MNQT}_c$($t_l$))/2 for $t_h$=30 and 15, respectively.}
 \label{fig:FRVar_NMOS_moving_avg_avgFR}
\end{figure*}

\begin{figure*}[ht]
\centering
  \includegraphics[scale=0.37]{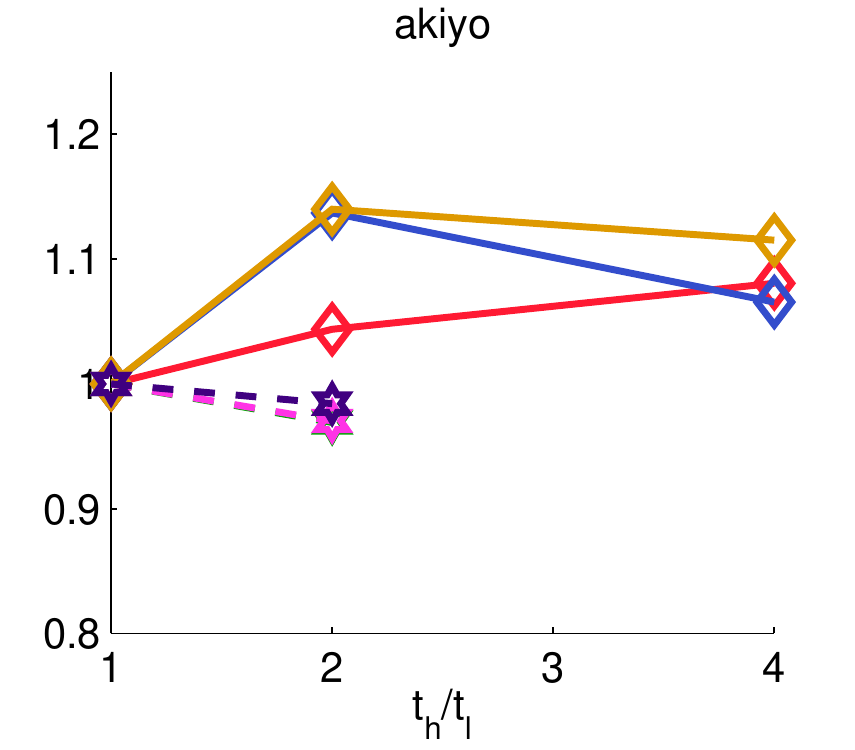}
  \includegraphics[scale=0.37]{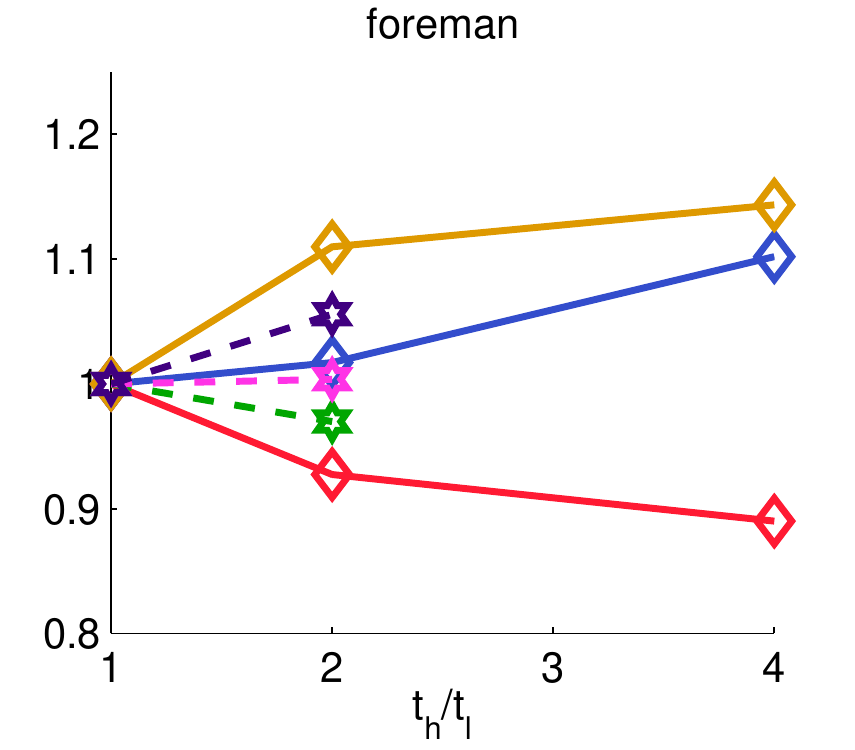}
  \includegraphics[scale=0.37]{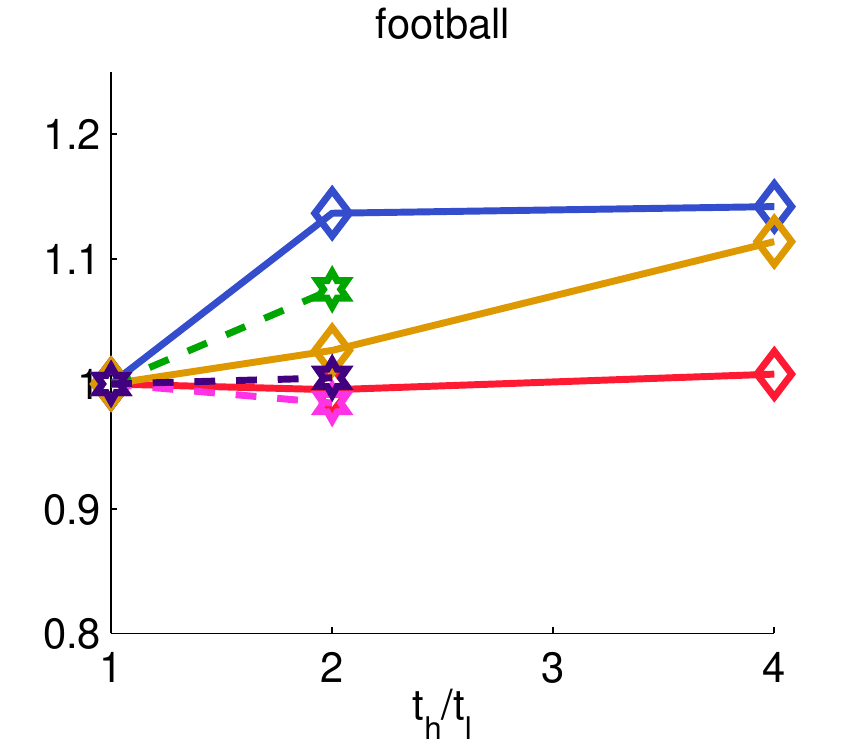}
  \includegraphics[scale=0.37]{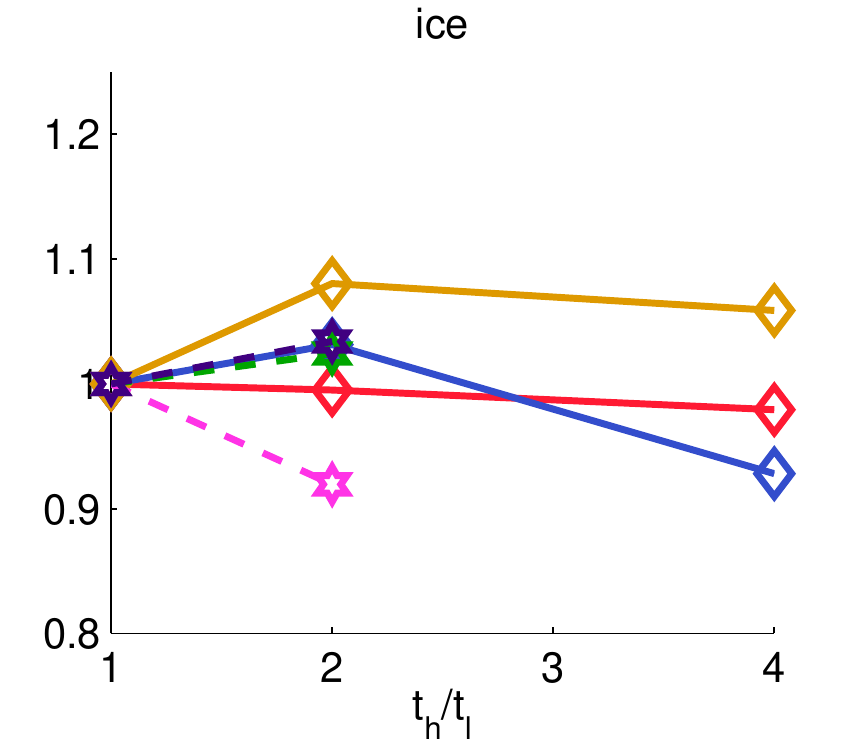}
  \includegraphics[scale=0.37]{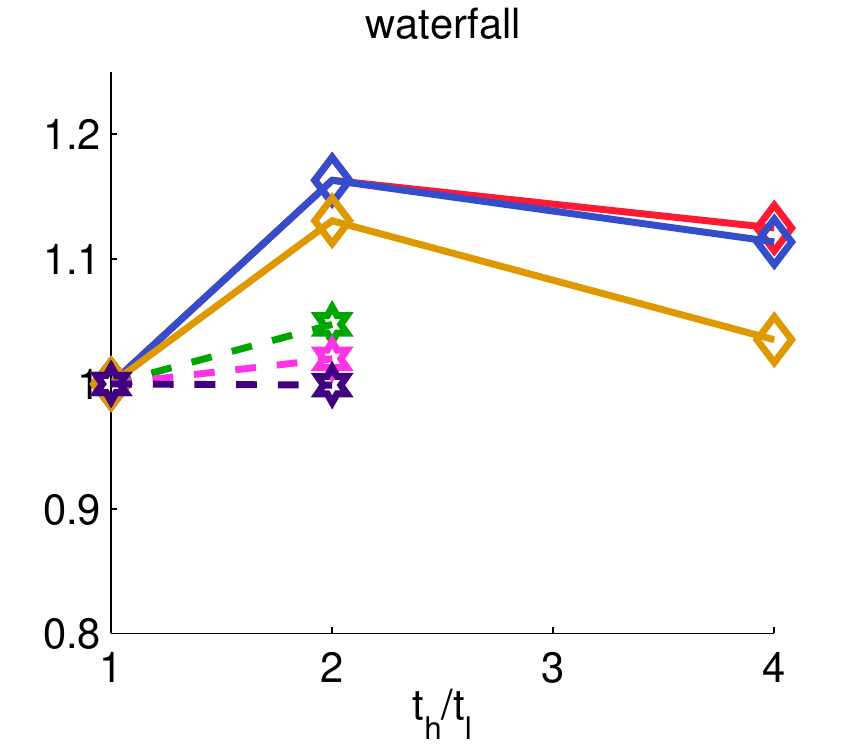}
  \includegraphics[scale=0.35]{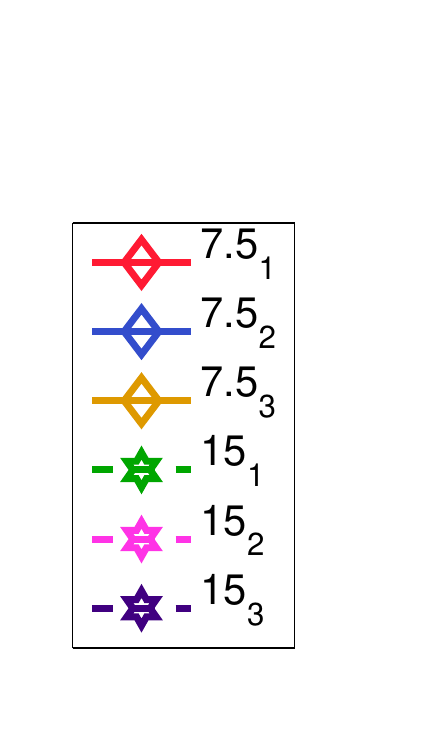}
 \caption{Q($t_h,t_l$)$/$Q($t_l,t_l$) vs. FR ratio ($t_h/t_l$) when $t_l$ is fixed. Lines with different markers and colors correspond to different $t_l$ and Fz, as indicated by the legends on the right. For example, the brown solid line as indicated by $7.5_3$ corresponds to data obtained with $t_l$=7.5, $t_h$=7.5, 15, 30, and Fz=3.}
 \label{fig:FRVar_NMOS_ratioFR_fl}
\end{figure*}

We use MOS($t_1$, $t_2$) to represent the mean opinion score (MOS) for a PVS with two oscillating frame rates $t_1$ and $t_2$.  In most practical applications, it is the relative MOS compared with some reference maximum MOS  that is of importance. Therefore, we report and model the relative quality ratings defined by Q($t_1,t_2$) = MOS($t_1,t_2$)/MOS($t_{\max}, t_{\max}$), with $t_{\max}$=30Hz.  Similarly, MOS($q_1, q_2$) denotes the MOS for a PVS with two oscillating QS's $q_1$ and $q_2$, and Q($q_1,q_2$) = MOS($q_1,q_2$)/MOS($q_{\min},q_{\min}$) with $q_{\min}$=16. The test video and test results (in terms of MOS) are available at~\cite{QPFRVar_dataset}.

\section{Video Quality under Frame Rate Variation}\label{sec:Test_results_FRVar}
This section considers the impact of periodic variation of FR on the perceptual quality, while the QS is fixed (QS=16). We first present results for videos with constant FR and show that the quality v.s. FR relation can be captured accurately by a model presented in~(\ref{eq:FRVar_MNQT_model}). We then report subjective ratings for videos with periodic FR variations under different variation magnitudes and frequencies. Based on these results, we further propose a model that relate the quality with both the low and high frame rates.

\subsection{Impact of Constant Frame Rate}\label{ssec:Imp_FR}
Figure ~\ref{fig:FRVar_pMOS_NQT_avgFR} shows Q($t_l, t_h$) v.s. $t/t_{\max}$ (here $t_{\max}$=30), of all the testing sequences with $t_l$=$t_h$=$t$. As expected, the MOS reduces as the frame rate decreases. In our prior work~\cite{yenfu_csvt} that examined the effect of FR and QS on the perceptual quality, when each is held constant over the entire test sequence, we have found that given a fixed QS, the impact of FR can be captured by the following model, known as model for normalized quality vs. temporal resolution (MNQT),
\begin{equation} \label{eq:FRVar_MNQT_model}
{\rm MNQT}_c(t) = \frac{1- e^{-\alpha_t \cdot(\frac{t}{t_{\max}})}}{1-e^{-\alpha_t}}.
\end{equation}
The model parameter $\alpha_t$ characterizes how fast the quality drops as the frame rate reduces with smaller value corresponding to faster dropping rate.
As can be seen from Fig.~\ref{fig:FRVar_pMOS_NQT_avgFR}, this model fits with the subset of measured data ($t_h=t_l$) very well with Pearson Correlation Coefficient (PCC)=$0.995$ and Root Mean Square Error (RMSE)=$0.013$. 

\begin{figure*}[ht]
\centering
  \includegraphics[scale=0.4]{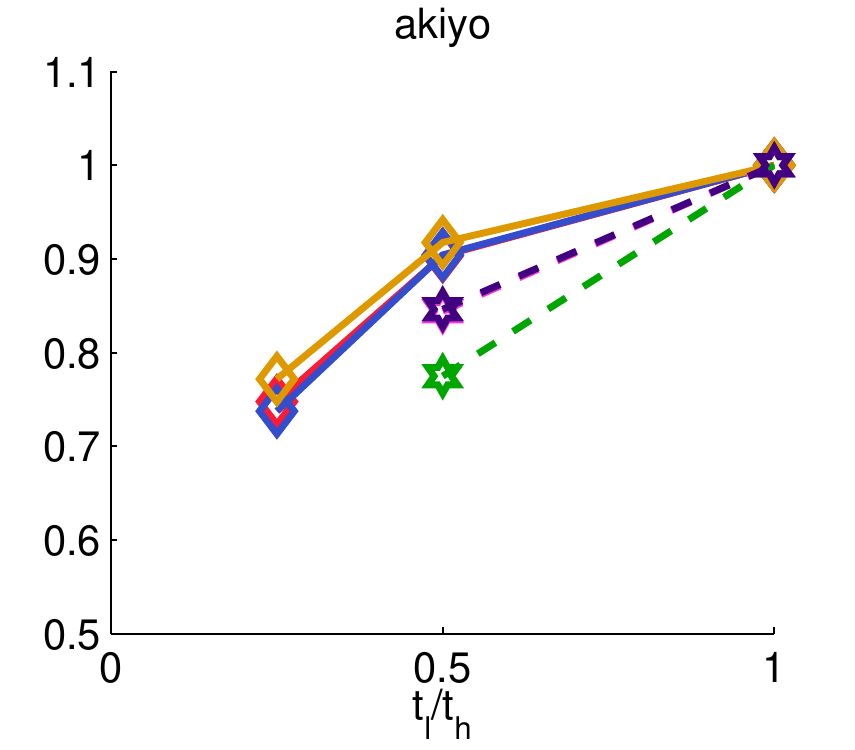}
  \includegraphics[scale=0.4]{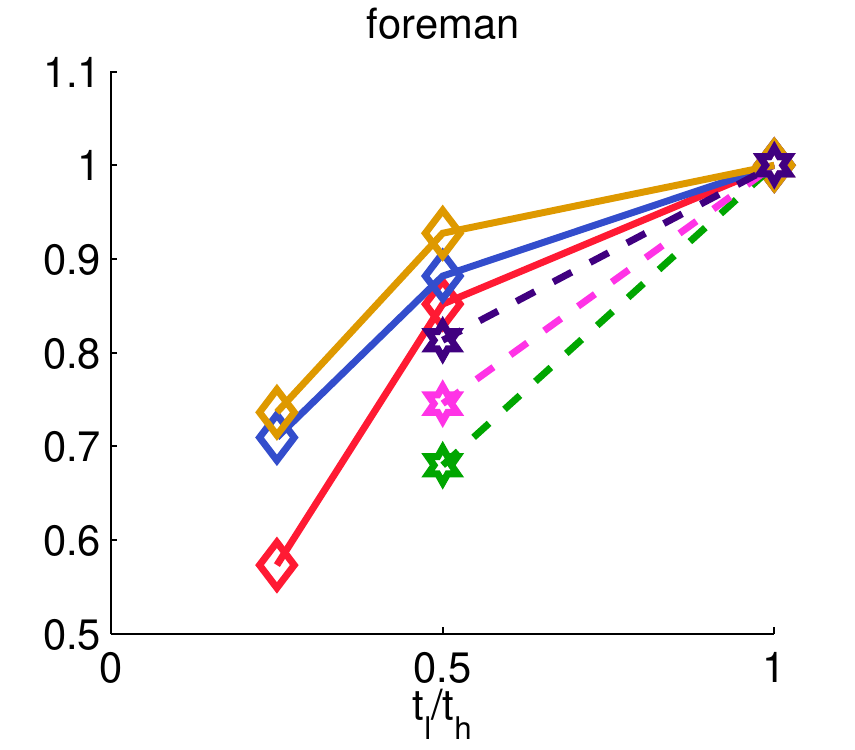}
  \includegraphics[scale=0.4]{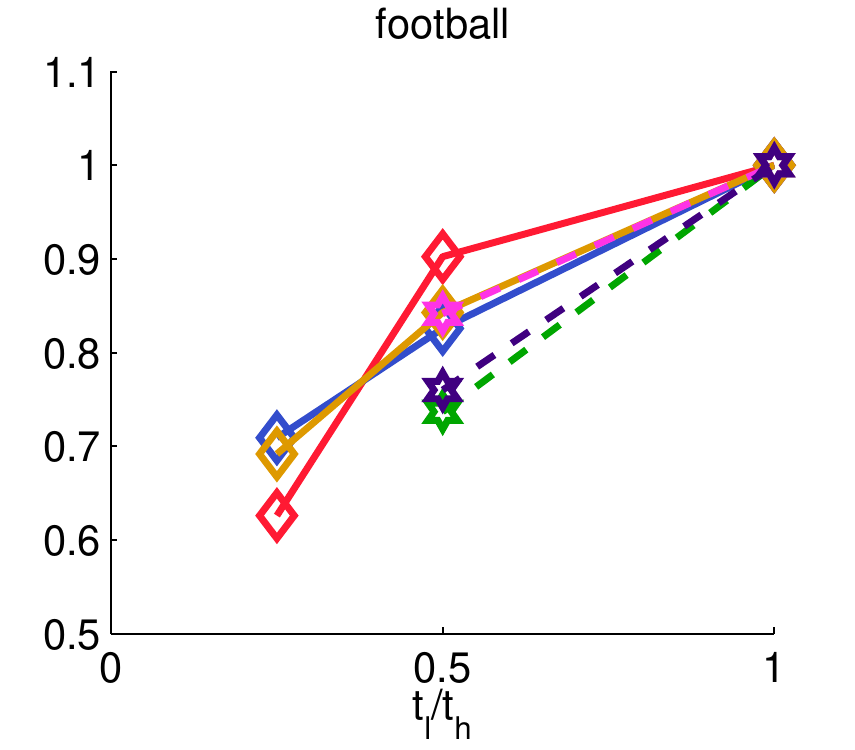}
  \includegraphics[scale=0.4]{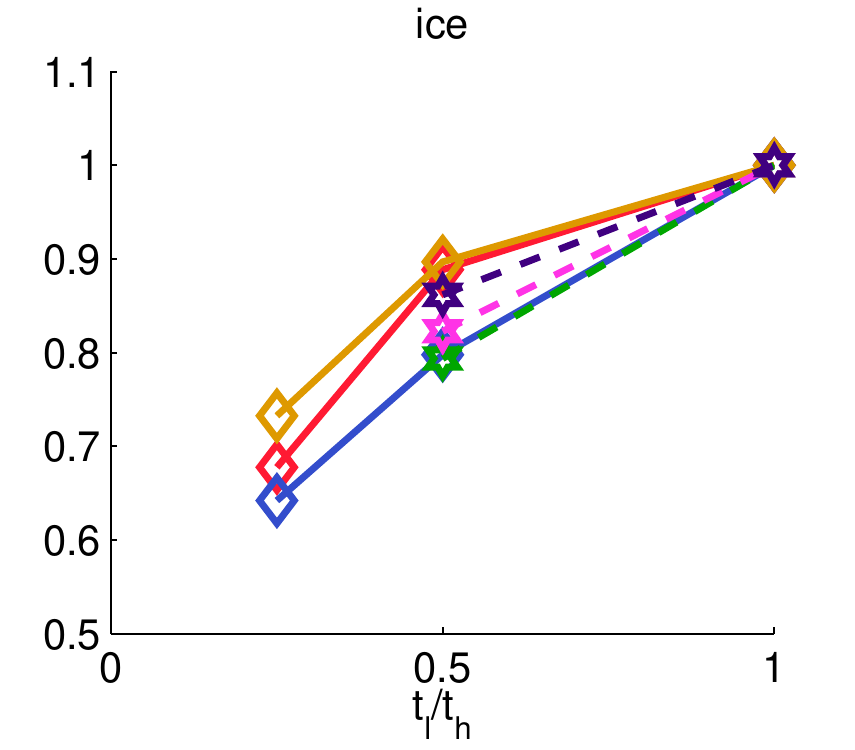}
  \includegraphics[scale=0.4]{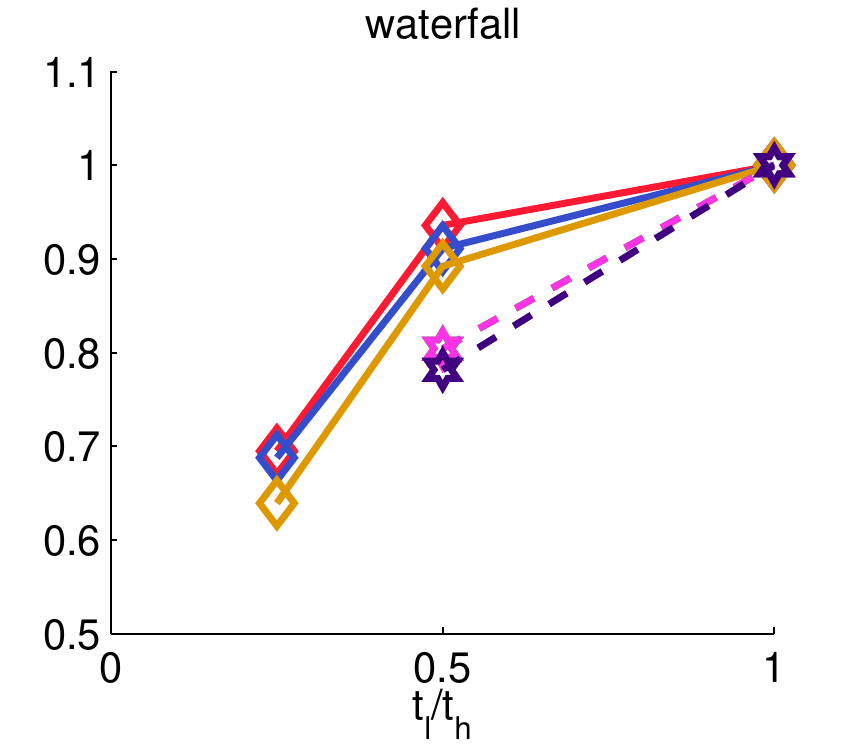}
 \caption{Q($t_h,t_l$)$/$Q($t_h,t_h$) vs. FR ratio ($t_l/t_h$)  when $t_h$ is fixed. Lines with different markers and colors correspond to different $t_h$ and Fz, as indicated by the legend in Fig.~\ref{fig:FRVar_NMOS_moving_avg_avgFR}}
 \label{fig:FRVar_NMOS_ratioFR_FR_h}
\end{figure*}

\begin{figure*}[ht]
\centering
  \includegraphics[scale=0.4]{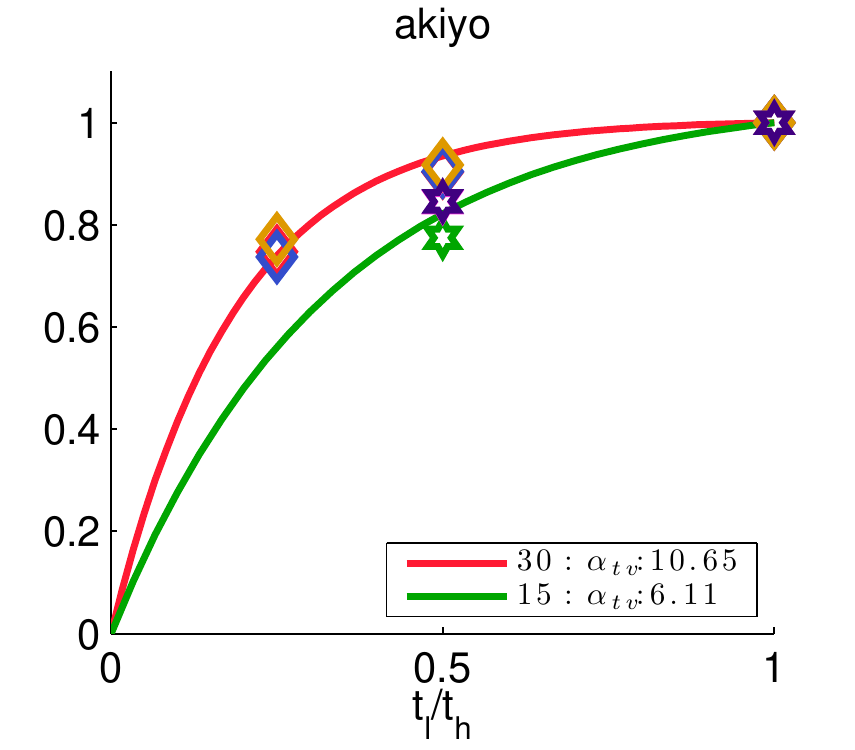}
  \includegraphics[scale=0.4]{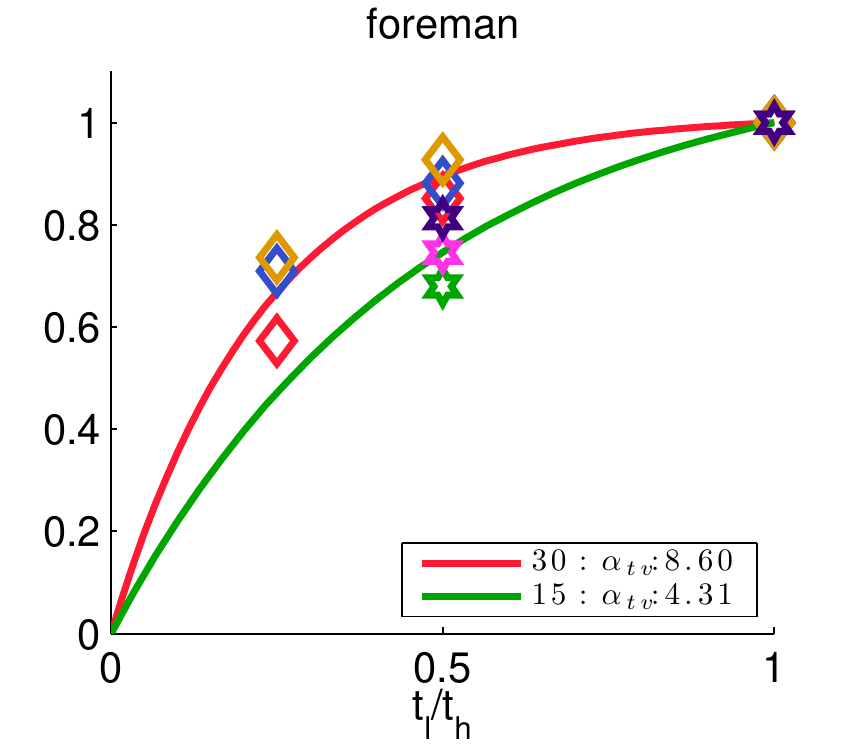}
  \includegraphics[scale=0.4]{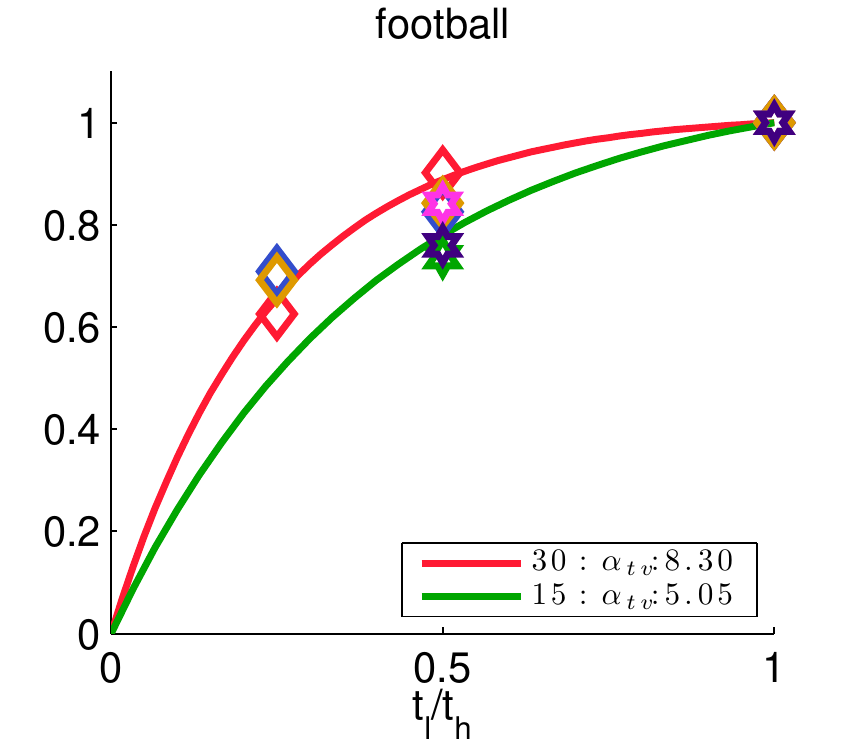}
  \includegraphics[scale=0.4]{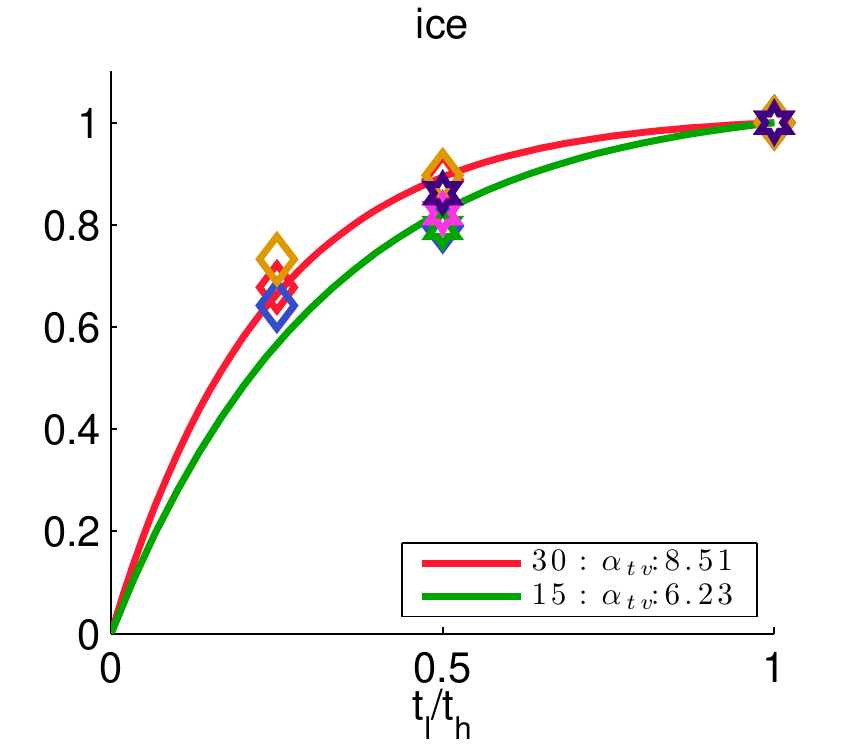}
  \includegraphics[scale=0.4]{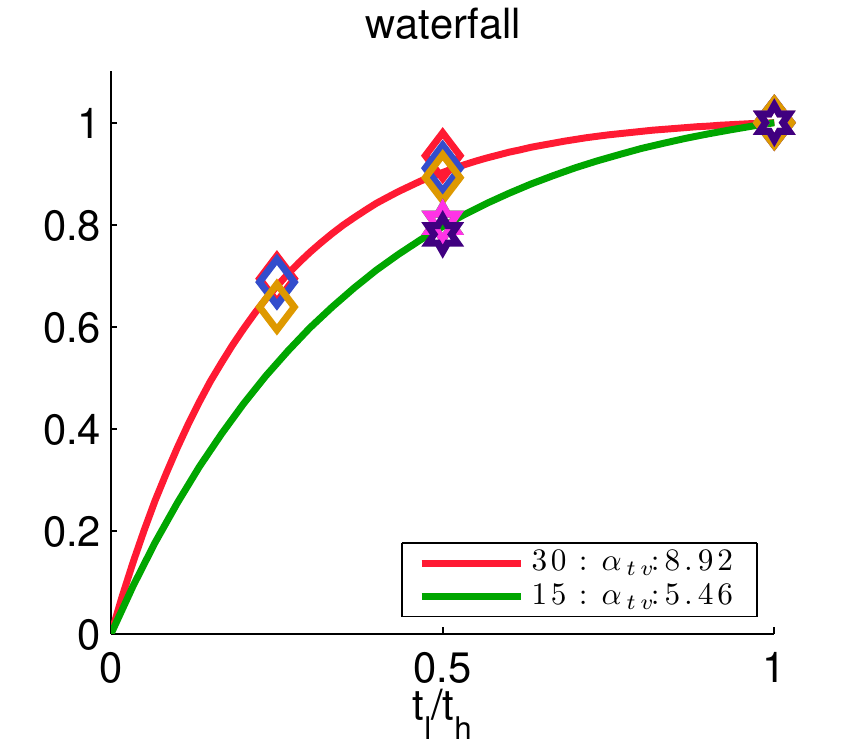}
 \caption{Predicted Q($t_h,t_l$)$/$Q($t_h,t_h$) vs. FR ratio ($t_l/t_h$) by an exponential model given in Eq.~(\ref{eq:FRVar_NMOS_ratioQS_invexp_model}) with PCC = $0.971$, RMSE = $0.029$. The model parameter $\alpha_{tv}$ depends on $t_h$, but not Fz.}\label{fig:FRVar_NMOS_ratioQS_invexp_2a1k}
\end{figure*}

\subsection{Impact of Frame Rate Variation}\label{ssec:Imp_FRVar}
We now consider sequences in which the frame rate alternates between $t_h$ and $t_l$. We first discuss, under the same average frame rate, $t_{avg}$ = ($t_h$+$t_l$)/2,  how does frame rate variation magnitude $\Delta t$=$t_h$-$t_l$ affects the perceived quality. Figure~\ref{fig:FRVar_NMOS_moving_avg_avgFR} shows that, when the $t_{avg}$ is the same, the MOS for a video with a constant frame rate (as indicated as cyan lines) is higher than that with frame rate variation. The degradation due to frame rate change is more severe when $\Delta t$ is higher (e.g., quality difference between (30,7.5) and constant frame rate of (30+7.5)/2=18.75 is greater than quality difference between (30,15) and constant frame rate of (30+15)/2=22.5.).  This result is as expected, as large frame rate variation induces noticeable jitter. It is interesting to note that points corresponding to $t_l$=$t_h$/2 are relatively close to the quality-frame rate curves achievable by using constant frame rates, for most of the sequences. But those with $t_l$ lower than $t_h$/2 are much below.

Figure~\ref{fig:FRVar_NMOS_moving_avg_avgFR} shows that using the quality model derived for constant FR, i.e., ${\rm MNQT}_c (t_{avg})$ with $t_{avg}$=($t_h$+$t_h$)/2,  to represent the quality of video with alternating FR's $t_h$ and $t_l$ is not very accurate. Another intuitive approach would be to model the perceived quality of a video with alternating FR between $t_h$ and $t_l$ by the average of the quality of a video with constant FR=$t_h$, predicted by ${\rm MNQT}_c(t_h)$, and that of a video with constant FR=$t_l$, predicted by ${\rm MNQT}_c(t_l)$. The results are also illustrated in Fig.~\ref{fig:FRVar_NMOS_moving_avg_avgFR}, where the navy solid lines and blue dotted lines are the predicted quality using this method when $t_h$ equals 30 and 15 Hz, respectively. The results indicate that this method is  more accurate than using ${\rm MNQT}_c(t_{avg})$, but is still not very accurate especially when $t_h$ is much larger than $t_l$.

To examine whether alternating between
$t_l$ and $t_h$ leads to better quality than staying at $t_l$, we plots Q($t_h, t_l$)/Q($t_l, t_l$) against the ratio $t_h/t_l$ in Figure~\ref{fig:FRVar_NMOS_ratioFR_fl}, where it shows that alternating between $t_l$ and $t_h$ generally leads to a quality that is better than or similar to that obtained by staying at $t_l$ (except for Foreman when $t_l$=7.5 and Fz=1, and for Ice when $t_l$=15 and Fz=2). The slope of improvement depends on Fz, $t_l$ and the source sequence.
However, the quality improvement becomes saturated as $t_h/t_l>$2. Furthermore, when $t_h>$2$t_l$, the quality reduces compared to when $t_h$=2$t_l$, for some cases, possibly due to the noticeable frame rate variation.

We next look at when $t_h$ is fixed, how the quality changes with different $t_l$. Figure~\ref{fig:FRVar_NMOS_ratioFR_FR_h} shows how Q($t_h, t_l$)/Q($t_h, t_h$) decreases with the frame rate ratio $t_l/t_h$.
We can see that the dropping trend depends on $t_h$, with a higher $t_h$ leading to a slower dropping rate. The dropping rate is also influenced by Fz, although the trend is inconsistent.

\begin{figure*}[ht]
\centering
  \includegraphics[scale=0.43]{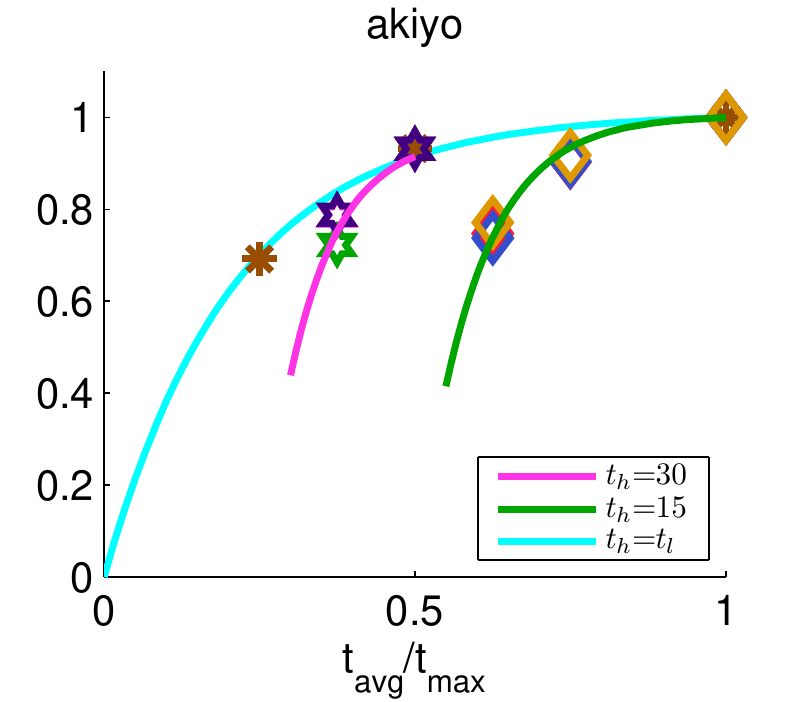}
  \includegraphics[scale=0.43]{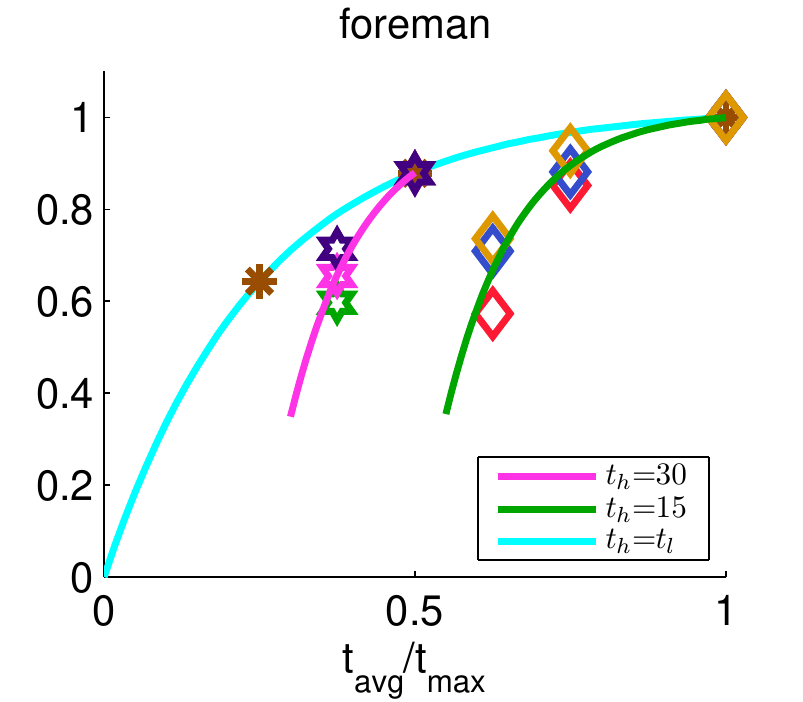}
  \includegraphics[scale=0.43]{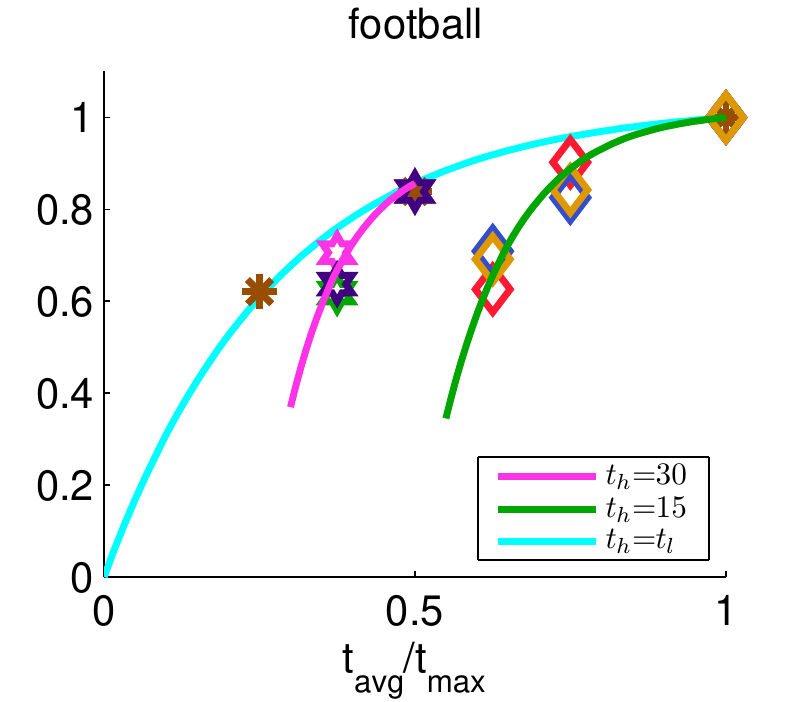}
  \includegraphics[scale=0.43]{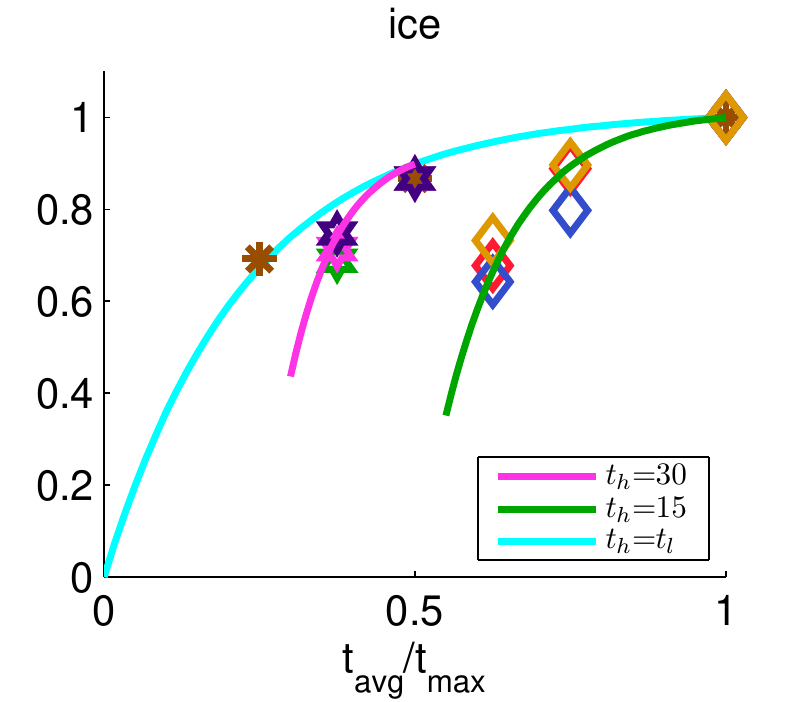}
  \includegraphics[scale=0.43]{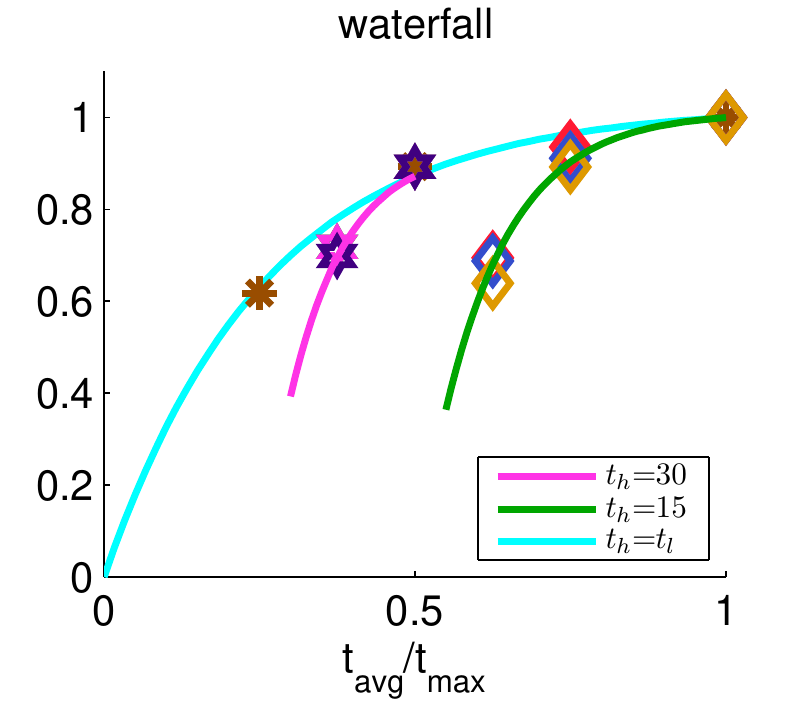}
 \caption{Predicted Q($t_h,t_l$) vs. $t_{avg}/t_{\max}$ using Eq.~(\ref{eq:FRVar_overallQ_invexp_invexp}) with PCC = $0.971$, RMSE = $0.029$.}\label{fig:FRVar_PrdNMOS_NFR_2a1k_invexp}
\end{figure*}

We have found that the quality drop with the frame rate ratio can be modeled well
using an inverse exponential function of the form
\begin{equation}\label{eq:FRVar_NMOS_ratioQS_invexp_model}
{\rm MNQT}_v(t_h,t_l) = \frac{1-e^{-\alpha_{tv}(t_h)\cdot(\frac{t_l}{t_h})}}{1-e^{-\alpha_{tv}(t_h)}},
\end{equation}
where the parameter $\alpha_{tv}$ depends on $t_h$. This is based on our earlier observation that the influence of Fz on the dropping rate is inconsistent, and also based on the ANOVA analysis described in Tab.~\ref{tab:FRVar_ANOVA_Fz} and Sec.~\ref{ssec:FRVar_ANOVA_test}. Fig.~\ref{fig:FRVar_NMOS_ratioQS_invexp_2a1k} presents the fitting curves using~(\ref{eq:FRVar_NMOS_ratioQS_invexp_model}).


\subsection{Overall Quality Model}
In order to predict the overall quality of a video with frame rates oscillating between $t_h$ and $t_l$, we make use of the fact that Q($t_h$, $t_l$) can be written as
\begin{equation}\label{eq:FRVar_def_form}
Q(t_h,t_l)=Q(t_h, t_h)\frac{Q(t_h,t_l)}{Q(t_h,t_h)}.
\end{equation}

Then we can use the model in~(\ref{eq:FRVar_MNQT_model}) to predict the first term and use the model in~(\ref{eq:FRVar_NMOS_ratioQS_invexp_model}) to estimate the second term. This yields the proposed quality model for videos with FR variation (to be called ${\rm QTV}$) given below
\begin{align}\label{eq:FRVar_overallQ}
{\rm QTV}(t_h,t_l) &= {\rm MNQT}_c(t_h) {\rm MNQT}_v(t_l, t_h).\\
&=\frac{1- e^{-\alpha_t \cdot(\frac{t}{t_{\max}})}}{1-e^{-\alpha_t}}\frac{1- e^{-\alpha_{tv}(t_h)\cdot(\frac{t_l}{t_h})}}{1-e^{-\alpha_{tv}(t_h)}}.\label{eq:FRVar_overallQ_invexp_invexp}
\end{align}

The first term in~(\ref{eq:FRVar_overallQ}) predicts the quality achievable with a constant high FR $t_h$, and the second term estimates the quality reduction when the FR fluctuates between $t_h$ and $t_l$.
Figure~\ref{fig:FRVar_PrdNMOS_NFR_2a1k_invexp} shows the model curves fit the measured data very well.
We note that these highly accurate results are obtained by finding best fitting model parameters for each sequence. Nonetheless, the fact that all sequences can be modeled well using the same model form is very encouraging. It suggests that the proposed simple model form reveals the right trend of how does $t_l$ and $t_h$ each affects the overall quality. Additional study is needed to investigate the relation between $\alpha_{tv}$ and $t_h$ and to develop a closed-form function possibly with content-dependent parameters. For the model to be useful for predicting quality of other sequences outside our test set, one must also investigate how to predict the model parameters from the video content.

\section{Video Quality under QS Variation}\label{sec:Test_results_QPVar}
This section presents the impact of periodic variation of QS on the perceptual quality, while the FR is fixed. We first demonstrate results for videos with constant QS and show that the quality v.s. QS relation can be approximated well by a model presented in~(\ref{eq:QPVar_MNQQ_model}). We then report subjective ratings for videos with periodic QS variations under different variation magnitudes and frequencies. Based on these results, we further propose a model that relate the quality with both the low and high QS.

\subsection{Impact of constant QS}\label{ssec:Imp_QP}
\begin{figure*}[ht]
\centering
  \includegraphics[scale=0.4]{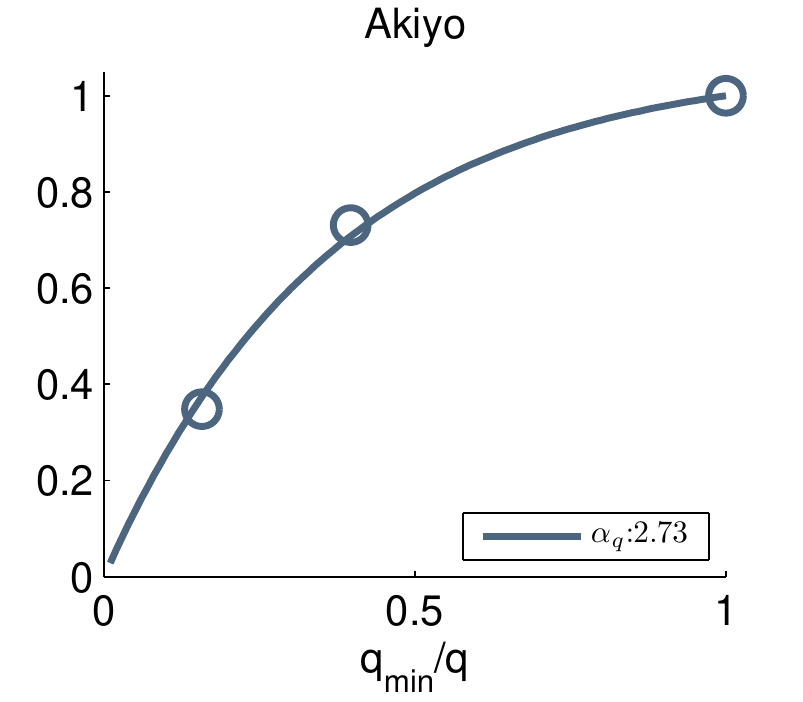}
  \includegraphics[scale=0.4]{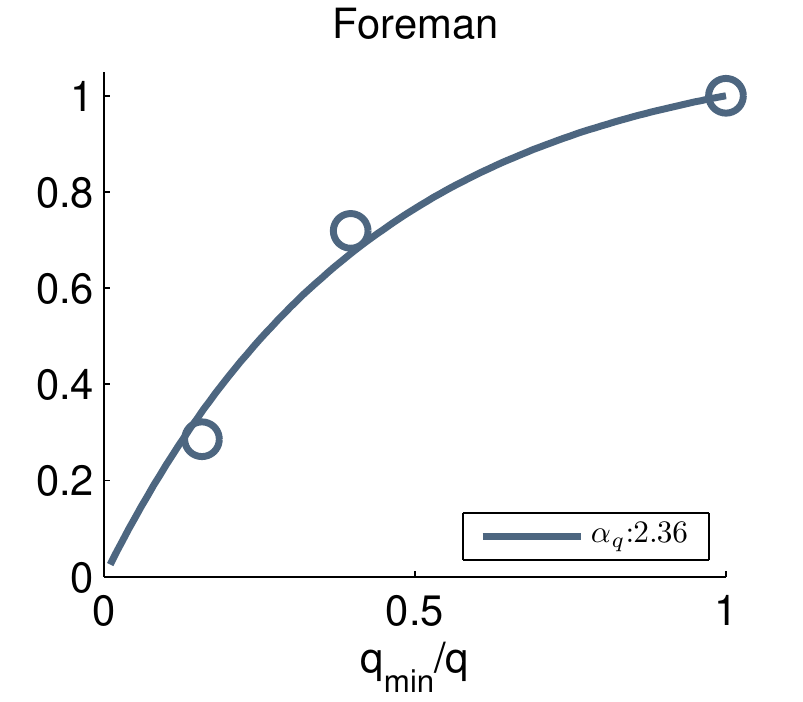}
  \includegraphics[scale=0.4]{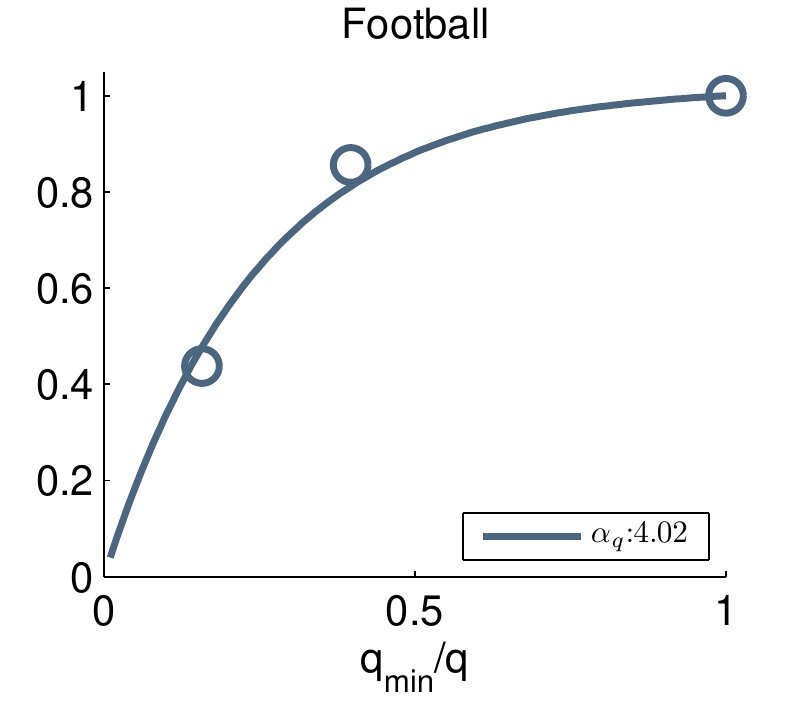}
  \includegraphics[scale=0.4]{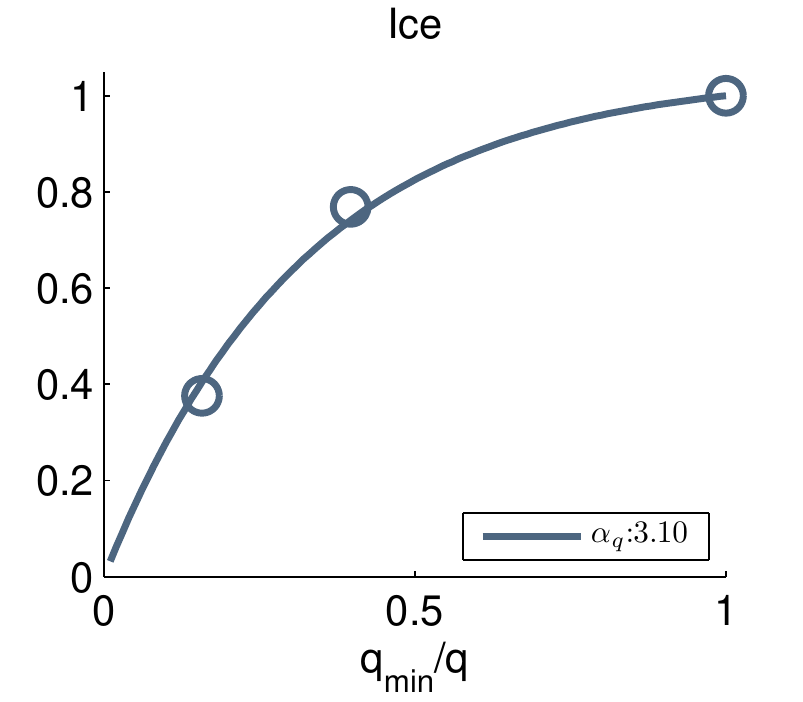}
  \includegraphics[scale=0.4]{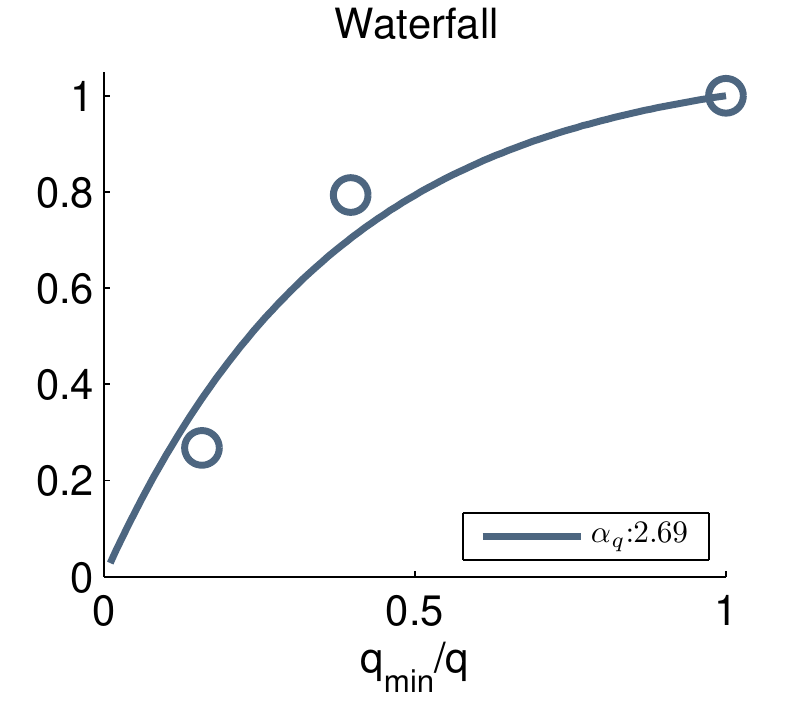}
 \caption{Constant QS case:Q($q_h, q_l$) vs. $q_{\min}/q$ for videos with $q_l=q_h=q$. Points are the measured data and curves are obtained using Eq.~(\ref{eq:QPVar_MNQQ_model}) with PCC=$0.990$, RMSE=$0.045$.}
 \label{fig:QPVar_MNQQ}
\end{figure*}

First we investigate the influence of the QS on the perceptual quality of a video when QS is constant. Figure~\ref{fig:QPVar_MNQQ} shows Q($q, q$) v.s. $q_{\min}/q$ for all the testing sequences with $q_h=q_l=q$. As expected, the quality reduces as $q$ increases (or $q_{min}/q$ decreases). In our prior work~\cite{yenfu_csvt}, we have examined the impact of QS on the video quality, when the FR and FS are fixed. There we have found that the impact of QS can be captured by the following model, known as model for normalized quality vs. quantization (MNQQ),
\begin{equation} \label{eq:QPVar_MNQQ_model}
{\rm MNQQ}_c(q) = \frac{1- e^{-\alpha_q (\frac{q_{\min}}{q})}}{1-e^{-\alpha_q}}.
\end{equation}
As can be seen from Fig.~\ref{fig:QPVar_MNQQ}, this model fits with the subset of measured data ($q_l=q_h$) very well, with PCC=$0.990$, RMSE=$0.045$.
Note that the parameter $\alpha_q$ characterizes how fast the quality drops as the QS increases, with a smaller value corresponding to faster drop.

%

%
\begin{figure*}[!htp]
\vspace{-.1in}
\centering
  \includegraphics[scale=0.37]{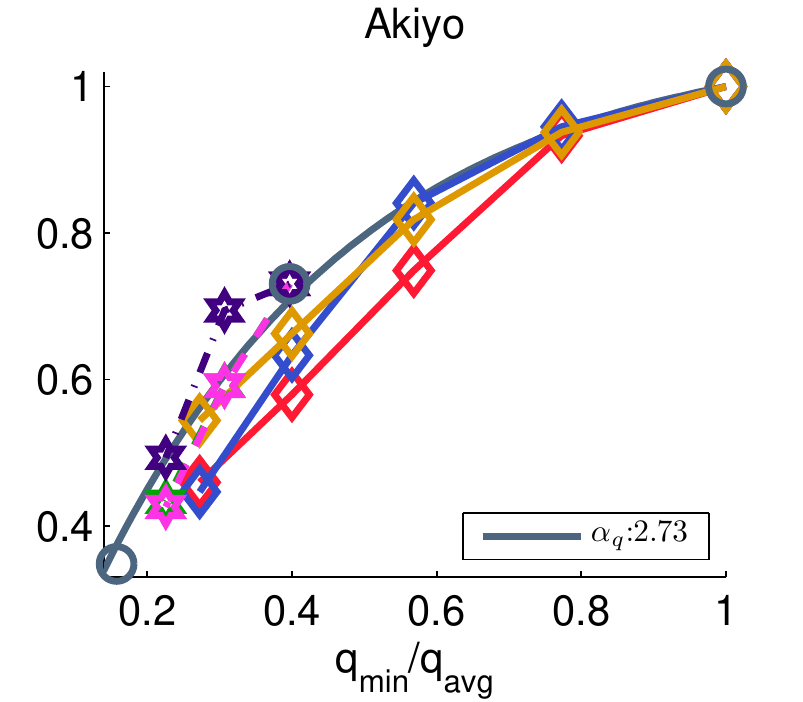}
  \includegraphics[scale=0.37]{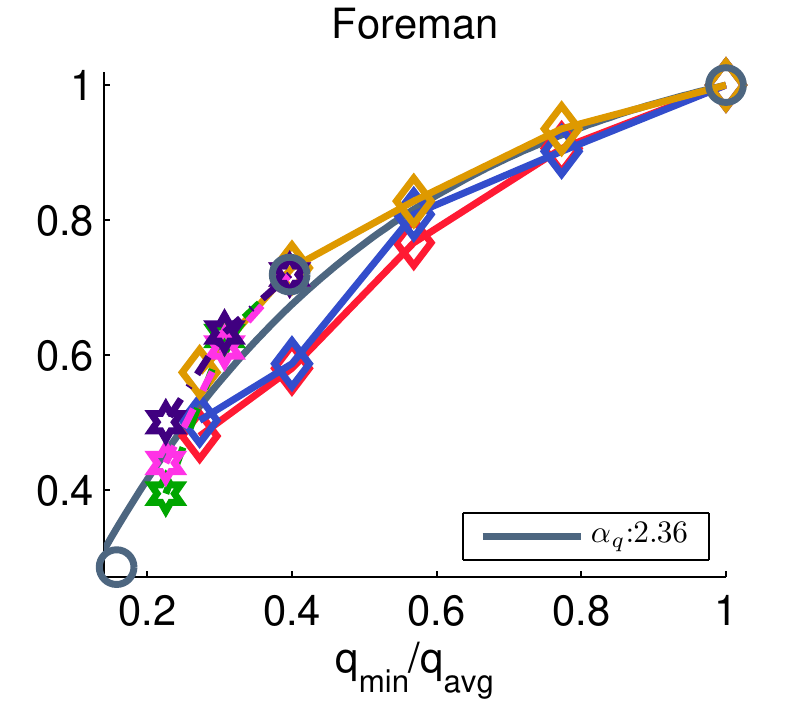}
  \includegraphics[scale=0.37]{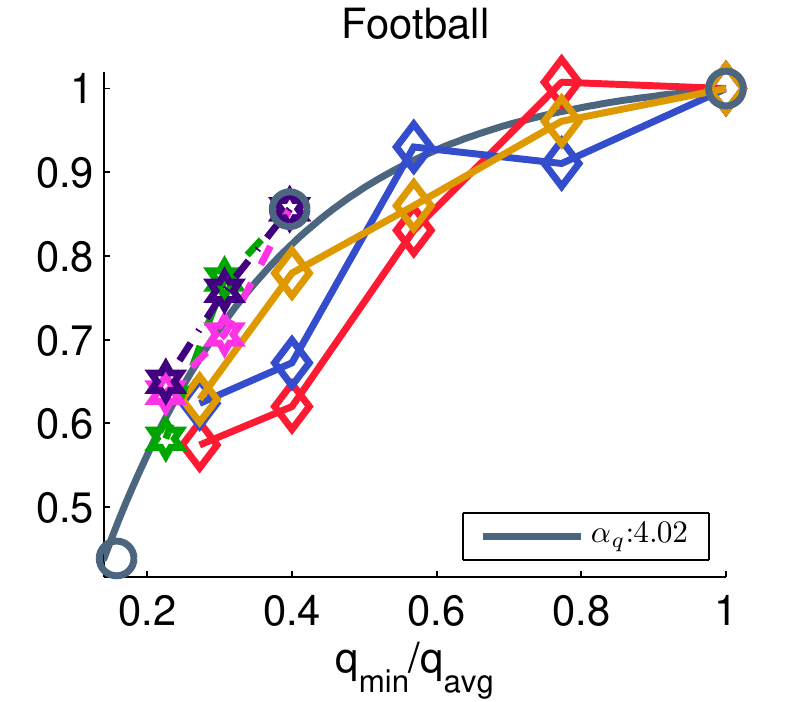}
  \includegraphics[scale=0.37]{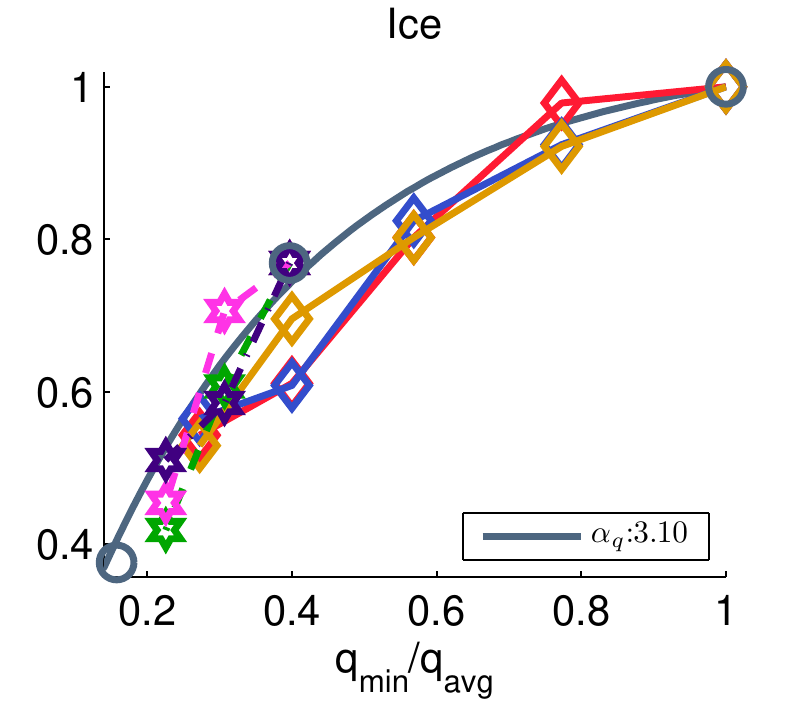}
  \includegraphics[scale=0.37]{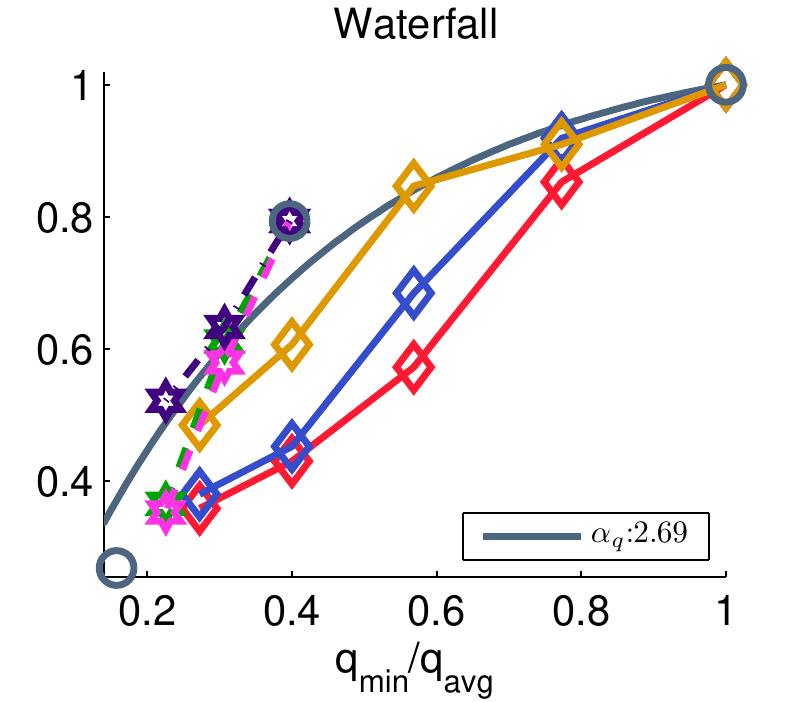}
  \includegraphics[scale=0.3]{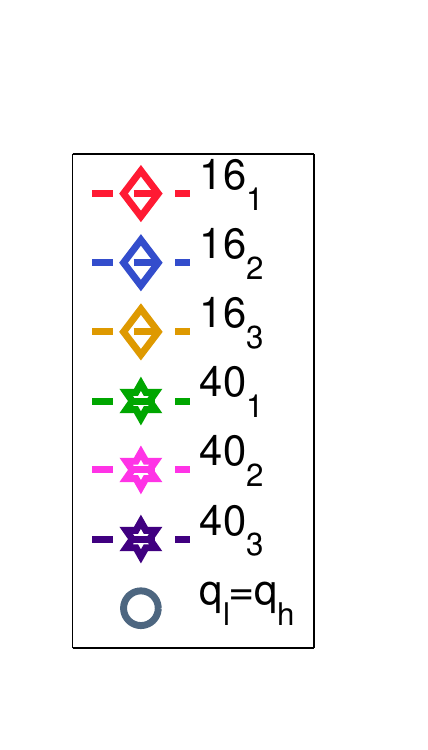}
 \caption{Q($q_l, q_h$) vs. $q_{\min}/q_{avg}$ when $q_l$ is fixed. Each line connects points with the same $q_l$ and Fz. For example, $16_3$ indicates $q_h$=16, Fz=3. The cyan and yellow dash curve are obtained from (${\rm MNQQ}_c$($q_l$=16)+${\rm MNQQ}_c$($q_h$))/2, and (${\rm MNQQ}_c$($q_l$=40)+${\rm MNQQ}_c$($q_h$))/2, respectively.}
 \label{fig:QPVar_MNQQ_moving_avg}
\end{figure*}
%
\begin{figure*}[ht]
\centering
  \includegraphics[scale=0.37]{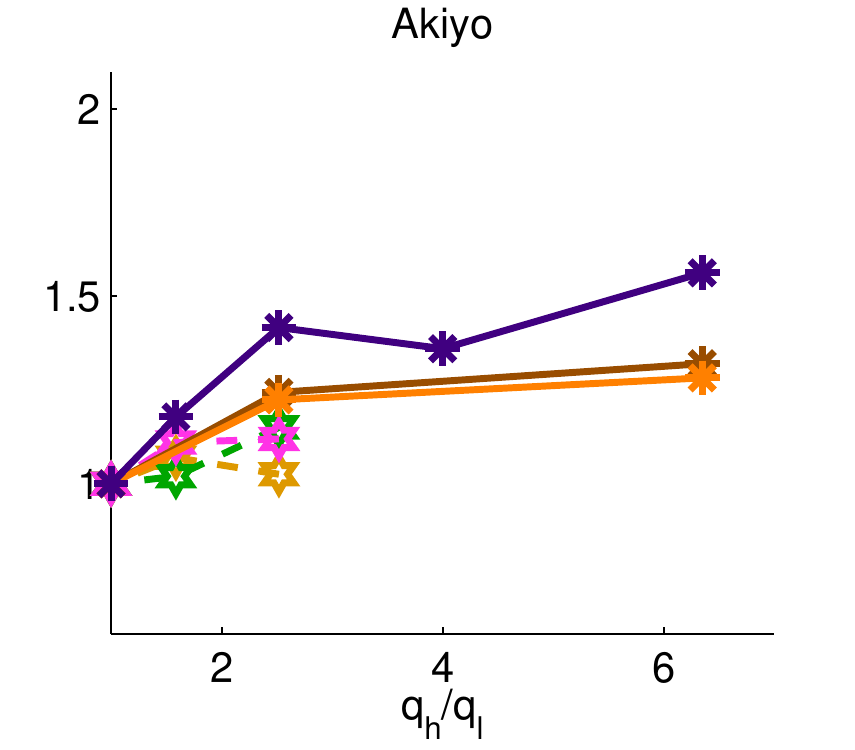}
  \includegraphics[scale=0.37]{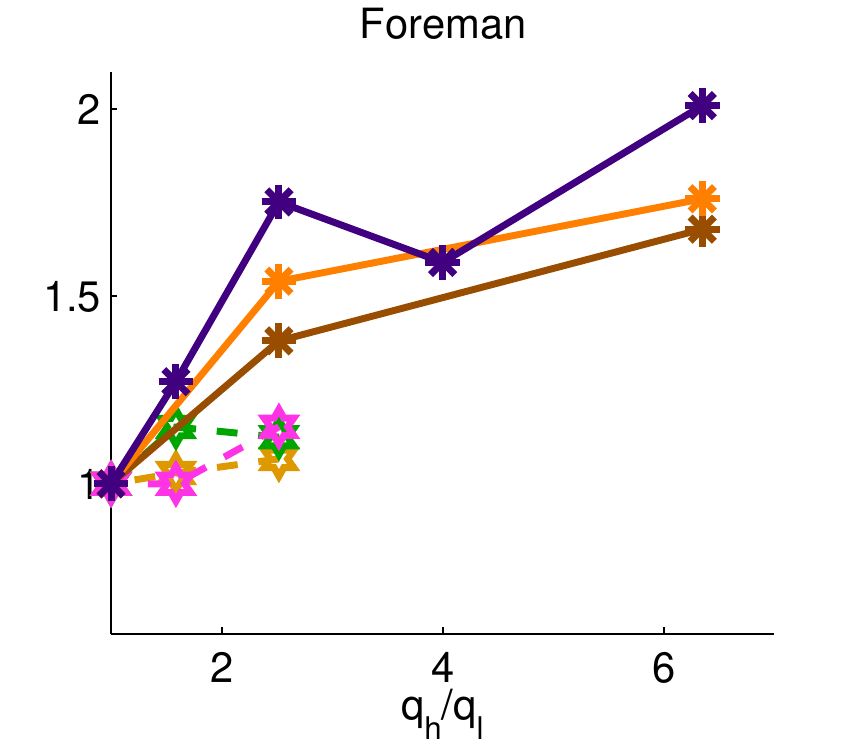}
  \includegraphics[scale=0.37]{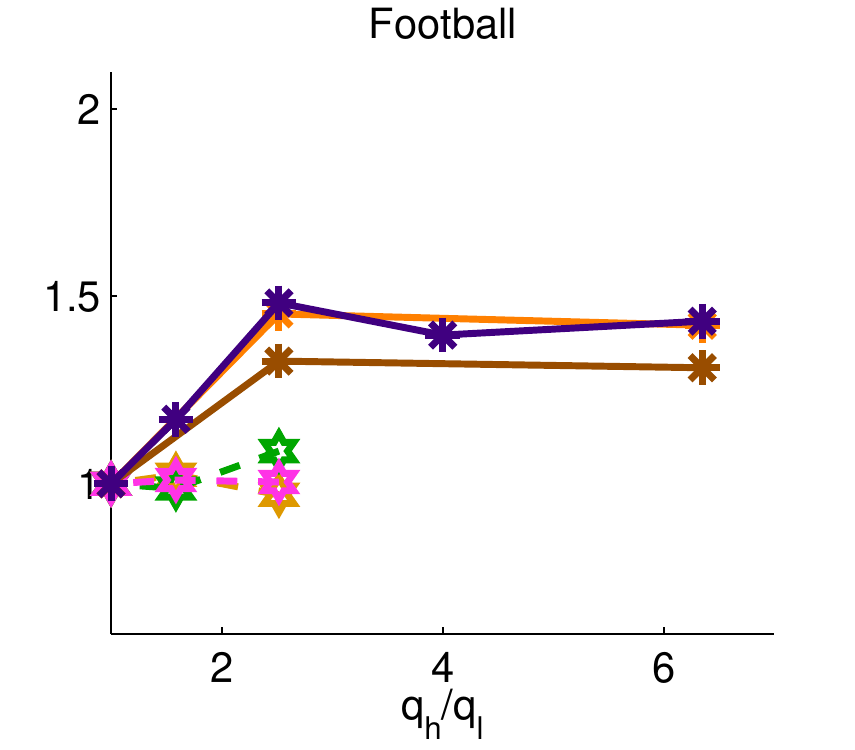}
  \includegraphics[scale=0.37]{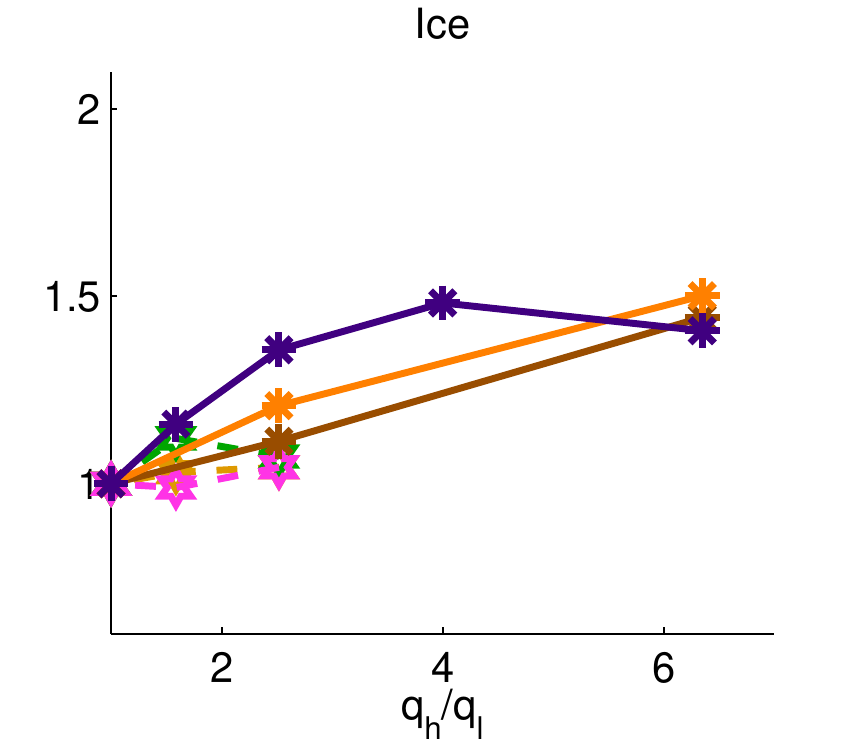}
  \includegraphics[scale=0.37]{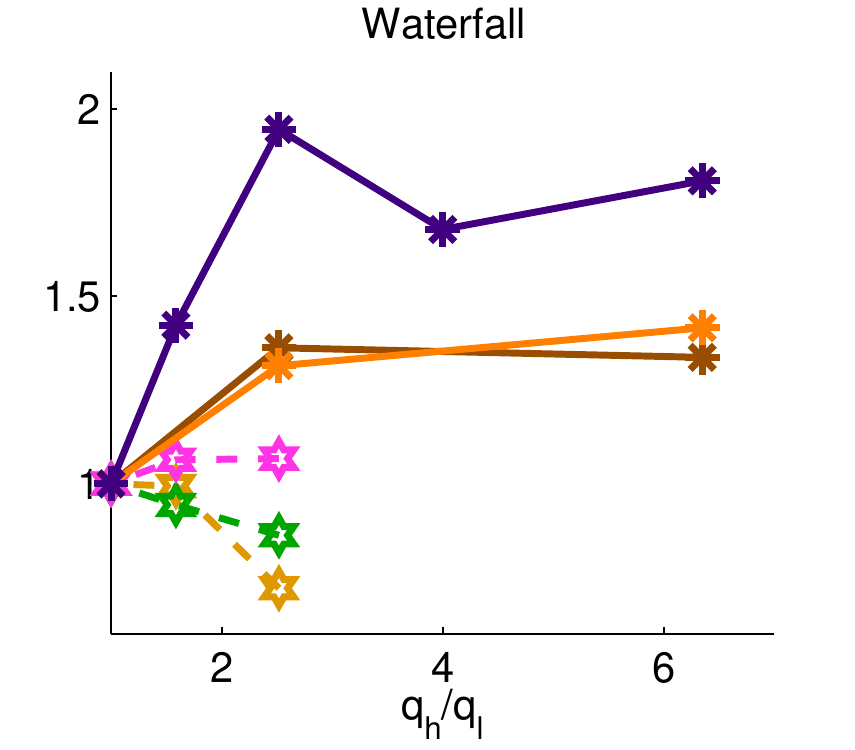}
  \includegraphics[scale=0.34]{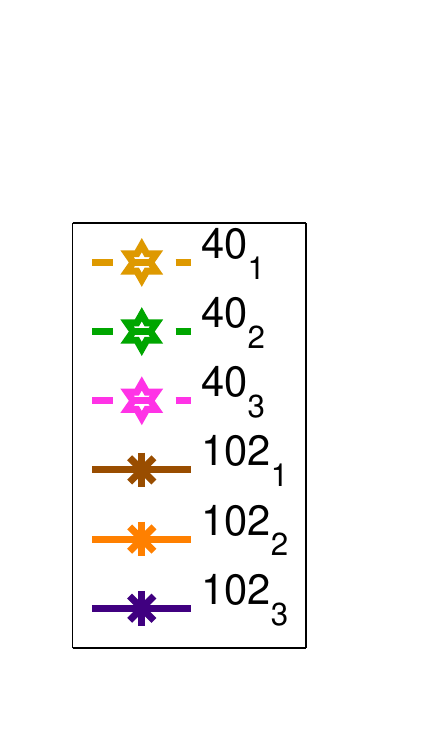}
 \caption{Q($q_h, q_l$)/Q($q_h, q_h$) vs. $q_l/q_h$ when $q_h$ is fixed. Each line connects points with the same $q_h$ and Fz. For example, $102_3$ indicates $q_h$=102, Fz=3.}
 \label{fig:NMOS_ql_qh_QP_h}
\end{figure*}

\subsection{Impact of QS Variation}\label{ssec:Imp_QPVar}
We now consider video clips in which the QS alternates between $q_h$ and $q_l$. We first discuss, under the same average QS, $q_{avg}$ = ($q_h$+$q_l$)/2,  how does QS variation affects the perceived quality. Figure~\ref{fig:QPVar_MNQQ_moving_avg} shows that, when $q_{avg}$ is the same, the quality for a video with a constant QS, Q($q_{avg}, q_{avg}$), is higher than that with QS variation, Q($q_h$, $q_l$). Note that the solid grayish blue curve in Fig.~\ref{fig:QPVar_MNQQ_moving_avg} are predicted quality for a video with constant QS based on (\ref{eq:QPVar_MNQQ_model}). It shows that using the quality model derived for constant QS, i.e., ${\rm MNQQ}_c (q_{avg})$  to represent the quality of video with alternating QS's $q_h$ and $q_l$ is not very accurate. Another natural attempt would be to model the perceived quality of a video with alternating QS between $q_l$ and $q_h$ by the average of the quality of a video with constant QS=$q_l$, predicted by ${\rm MNQQ}_c(q_l)$, and that of a video with constant QS=$q_h$, predicted by ${\rm MNQQ}_c(q_h)$. The result is also included in Fig.~\ref{fig:QPVar_MNQQ_moving_avg}; the cyan solid lines and yellow dash lines indicate the predicted quality using this method when $q_l$ equals 16 and 40, respectively. It can be seen that the prediction is not accurate especially when $q_l$ is low, e.g., 16.

We next examine when $q_h$ is fixed due to the lowest available bandwidth, whether alternating between $q_l$ and $q_h$ leads to better quality than staying at $q_h$. We plot Q($q_l, q_h$)/Q($q_h, q_h$) against the ratio $q_h/q_l$ in Fig.~\ref{fig:NMOS_ql_qh_QP_h}. For $q_h$=102, alternating between $q_l$ and $q_h$, is consistently better than staying at $q_h$=102, and the degree of improvement depends on Fz and the texture details of the video (e.g., Waterfall and Foreman have higher improvement ratio, e.g., up to 2). It can be observed that shorter Fz leads to less improvement. This is as expected as shorter Fz corresponds to more rapid QS switching, which can be more annoying to the human viewers. Interestingly, the slope of improvement saturates and becomes inconsistent as $q_l$ further decreases as $q_h/q_l$ becomes higher than 2.5. When $q_h$ is already low (e.g. $q_h$=40), switching to $q_l$ provides inconsistent gain for most sequences. For Waterfall, we see a consistent quality drop when Fz=1 and 2. Our ANOVA analysis (described in Sec.~\ref{sec:ANOVA_test}) shows that the quality variation observed for both $q_l$=25 and $q_l$=16 is statistically insignificant (see entries for ${\rm P_2}$ and ${\rm P_3}$ in Tab.~\ref{tab:ANOVA_QPVar}).

\begin{figure*}[ht]
\centering
  \includegraphics[scale=0.4]{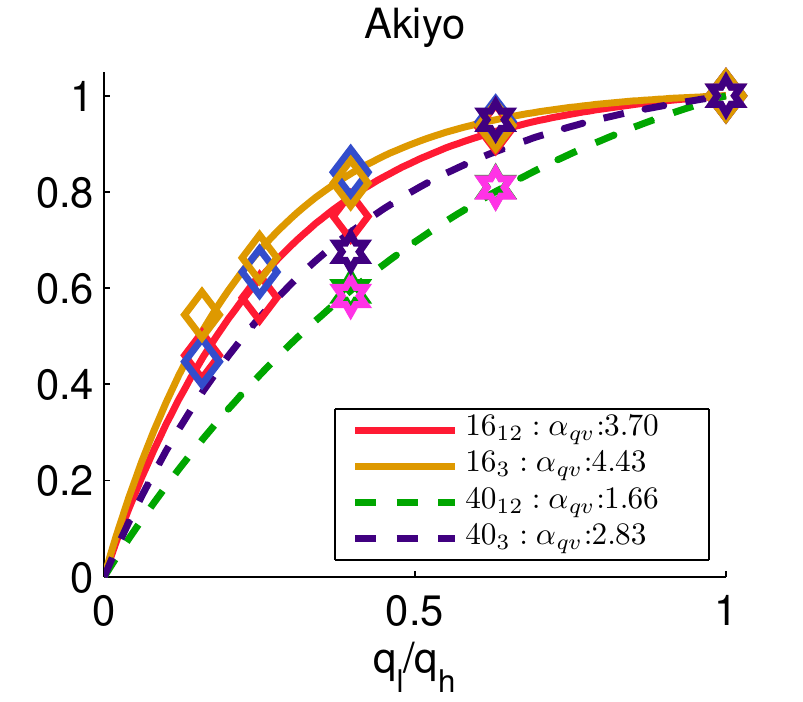}
  \includegraphics[scale=0.4]{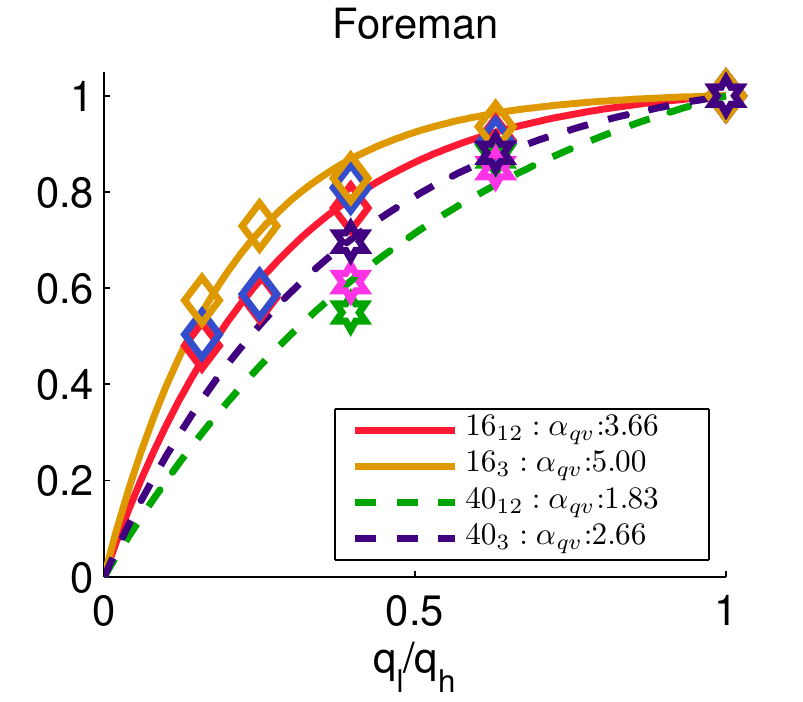}
  \includegraphics[scale=0.4]{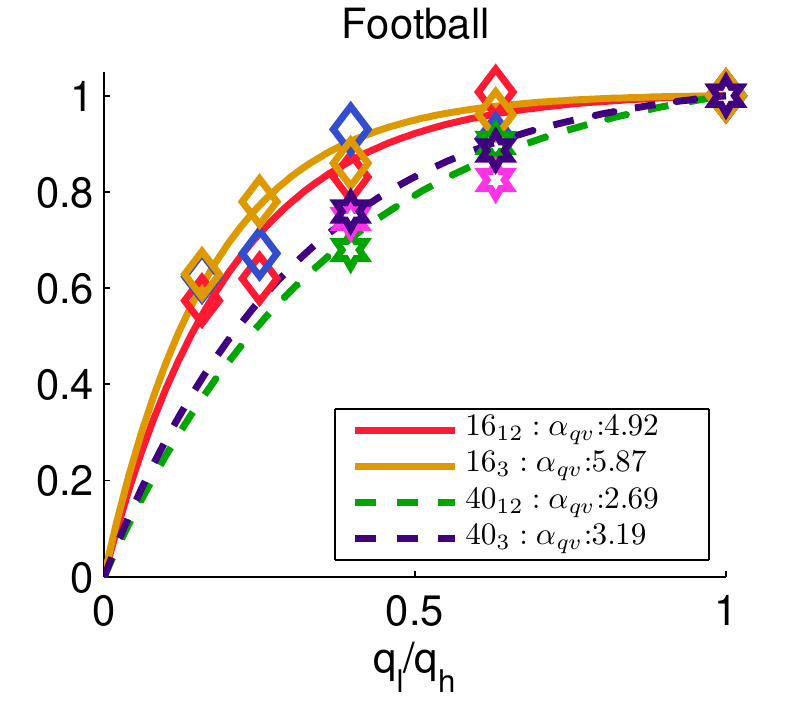}
  \includegraphics[scale=0.4]{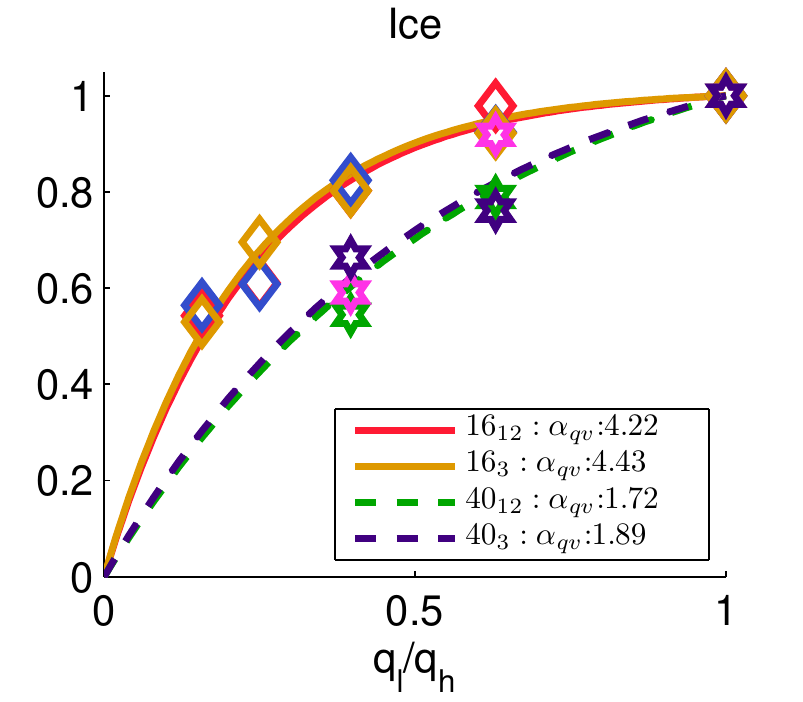}
  \includegraphics[scale=0.4]{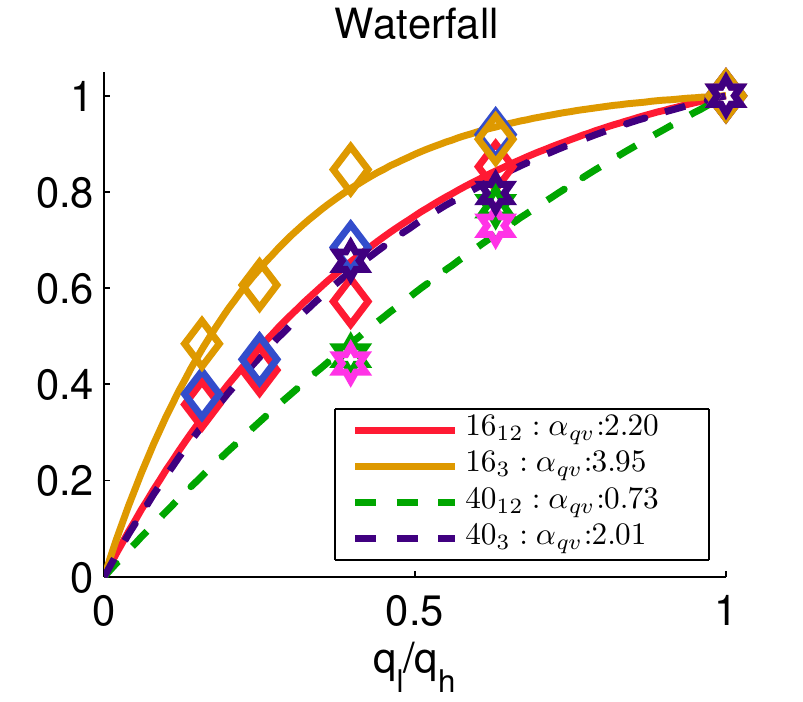}
 \caption{$Q(q_h,q_l)/Q(q_l, q_l)$ vs. QS ratio ($q_l/q_h$). The model curves are based on the model in (\ref{eq:QPVar_invexp}) with PCC = $0.983$, RMSE = $0.035$. $\alpha_{qv}$ depends on $t_h$ and Fz but is the same for Fz=1 and 2.}\label{fig:QPVar_NMOS_ratioQS_invexp_4par}
\end{figure*}

Next, we examine how the quality reduces when the QS varies between $q_l$ and $q_h$, compared to the quality achievable with a constant QS=$q_l$. Figure~\ref{fig:QPVar_NMOS_ratioQS_invexp_4par} plots Q($q_l, q_h$)/Q($q_l, q_l$) v.s. $q_l/q_h$. Here we can see that the falling rate depends on the video content and $q_l$. The dropping rate is very similar for Fz=1 and Fz=2,  but is typically slower when Fz=3.

We have found that Q($q_l, q_h$)/Q($q_h, q_h$) can be modeled quite accurately by the inverted exponential function of the QS ratio $q_l/q_h$, i.e.,
\begin{align}\label{eq:QPVar_invexp}
{\rm MNQQ}_v(q_h,q_l) &= \frac{1- e^{-\alpha_{qv}({\rm Fz}, q_l)\cdot(\frac{q_l}{q_h})}}{1-e^{-\alpha_{qv}({\rm Fz}, q_l)}},
\end{align}
where $\alpha_{qv}$ is a model parameter reflecting the dropping rate.
Based on the observations noted previously, and the ANOVA analysis described in Sec.~\ref{ssec:QPVar_ANOVA_test} and Tab.~\ref{tab:QPVar_ANOVA_Fz}, we choose to use the same model parameter $\alpha_{qv}$ for Fz=1 and 2, but a different one for Fz=3. Figure.~\ref{fig:QPVar_NMOS_ratioQS_invexp_4par} shows that the model predicts the measured data very well.

\subsection{Overall Video Quality}
In order to predict the overall quality of a video with alternating QS, we again recognize that Q($q_h,q_l$) can be written as
\begin{equation}\label{eq:QPVar_def_form}
Q(q_h,q_l)=Q(q_l, q_l)\frac{Q(q_h,q_l)}{Q(q_l,q_l)}
\end{equation}

\begin{figure*}[ht]
\centering
  \includegraphics[scale=0.37]{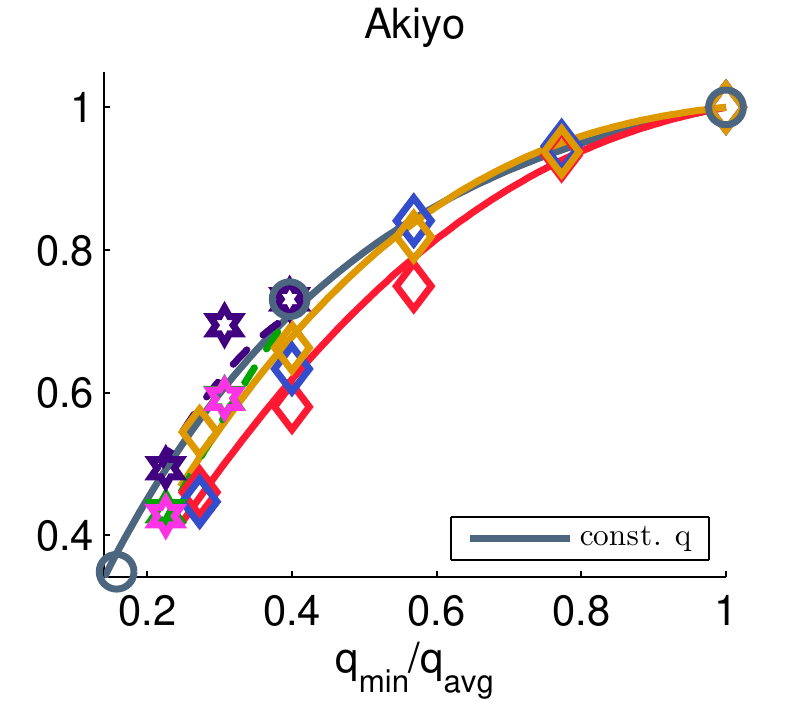}
  \includegraphics[scale=0.37]{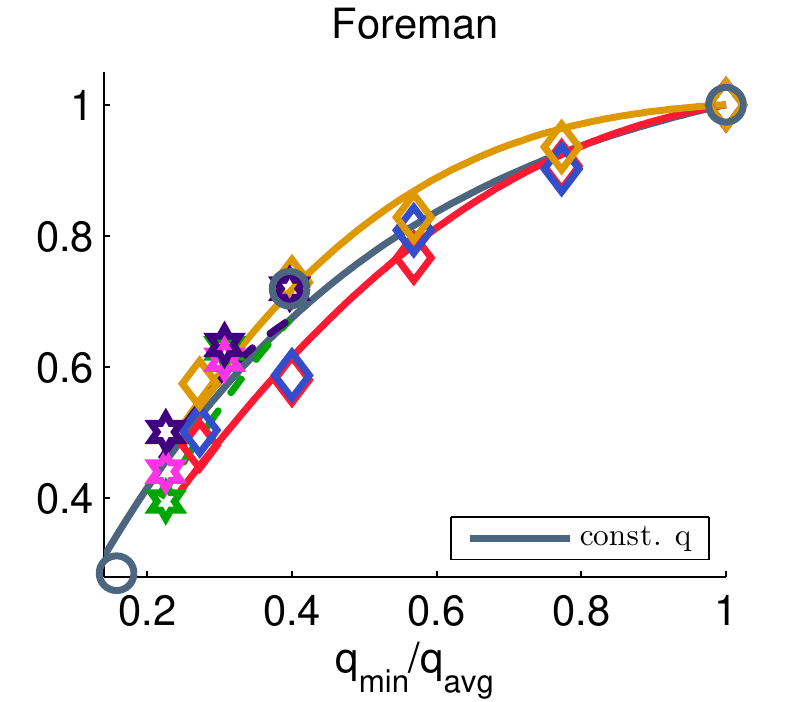}
  \includegraphics[scale=0.37]{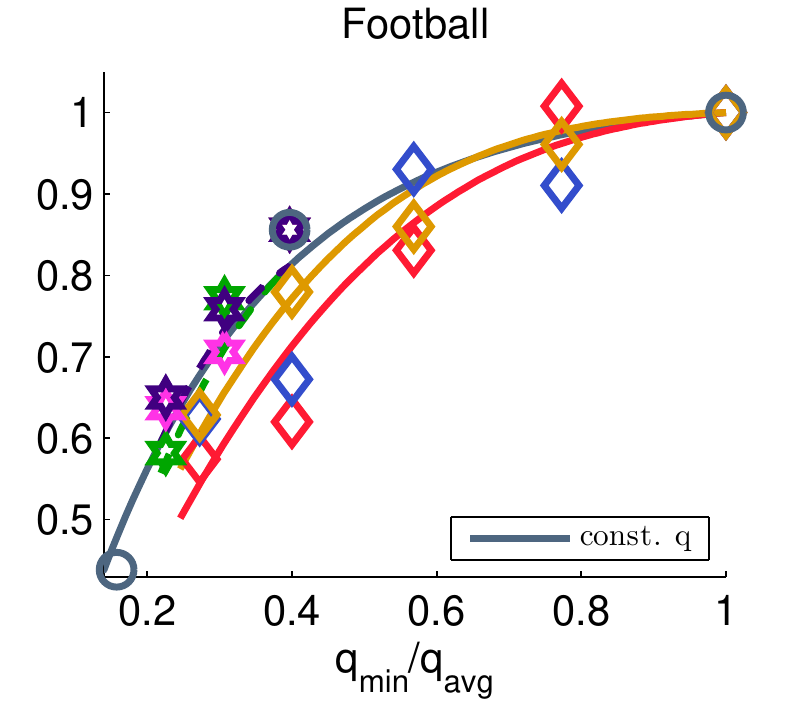}
  \includegraphics[scale=0.37]{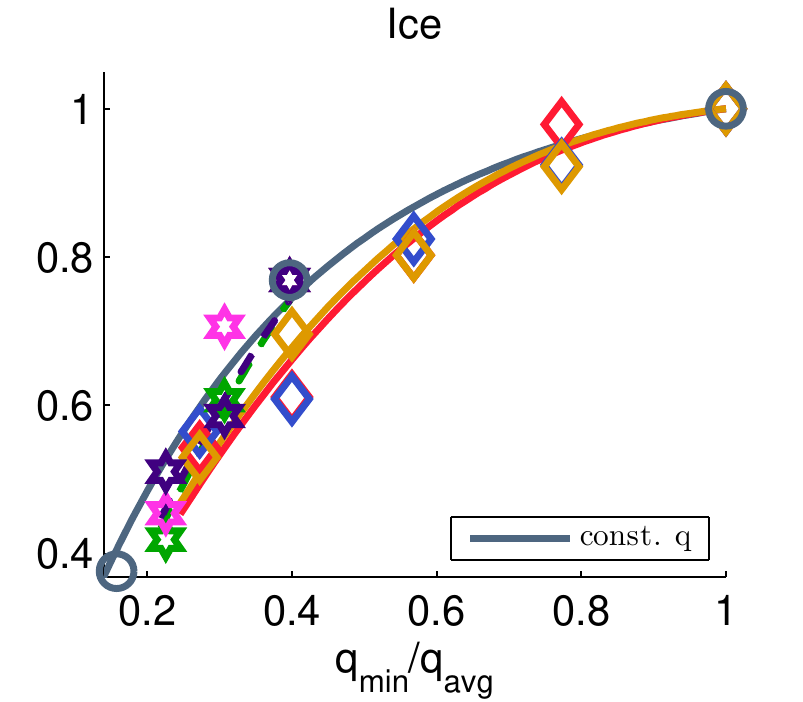}
  \includegraphics[scale=0.37]{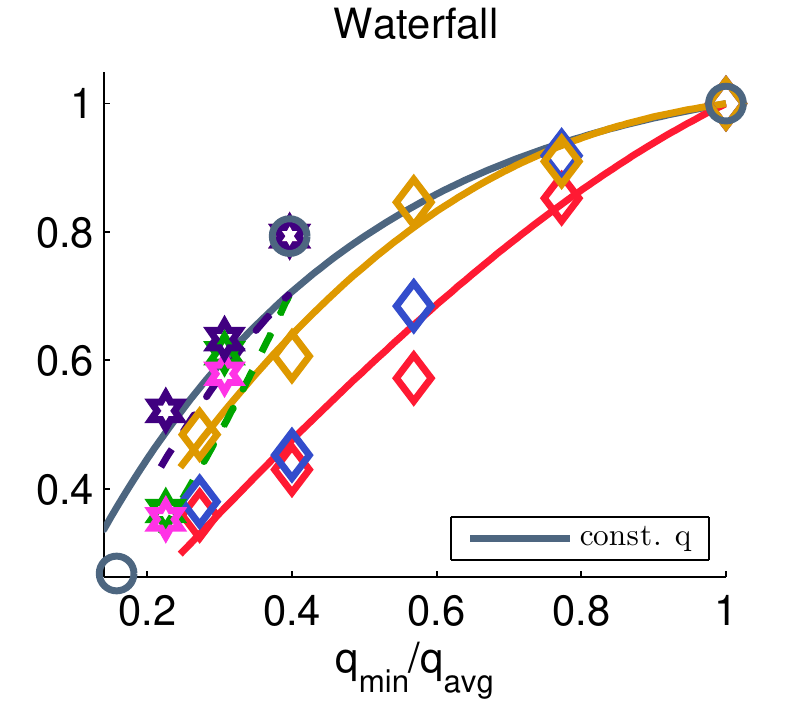}
  \includegraphics[scale=0.34]{QPVar_qql_Legend.pdf}
\caption{Q($q_h, q_l$) vs. $q_{\min}/q_{avg}$. Points are measured data and curves are based on the model in~(\ref{eq:overallQ_invexp_invexp}) with PCC = $0.978$, RMSE = $0.042$.}\label{fig:QPVar_NMOS_NQS_invexp_2par}
\end{figure*}

Then we can use the model in~(\ref{eq:QPVar_MNQQ_model}) to predict the term and use the model in~(\ref{eq:QPVar_invexp}) to approximate the second term. This yields the proposed quality model for videos with QS variation (to be called ${\rm QQV}$) given below
\begin{align}\label{eq:QPVar_overallQ}
{\rm QQV}(q_h,q_l) &= {\rm MNQQ}_c(q_l) {\rm MNQQ}_v(q_l, q_h),\\
&=\frac{1- e^{-\alpha_q (\frac{q_{\min}}{q_l})}}{1-e^{-\alpha_q}} \frac{1- e^{-\alpha_{qv}({\rm Fz}, q_l)\cdot(\frac{q_l}{q_h})}}{1-e^{-\alpha_{qv}({\rm Fz}, q_l)}}.\label{eq:overallQ_invexp_invexp}
\end{align}
Figure~\ref{fig:QPVar_NMOS_NQS_invexp_2par} illustrates that the predicted quality fits with the measured data very well. Note that the first term in~(\ref{eq:QPVar_MNQQ_model}) is responsible for predicting the quality at a constant QS $q_l$, whereas the second term accounts for the quality degradation due to QS variation.
%

\begin{figure*}[!ht]
\centering
  \includegraphics[scale=0.41]{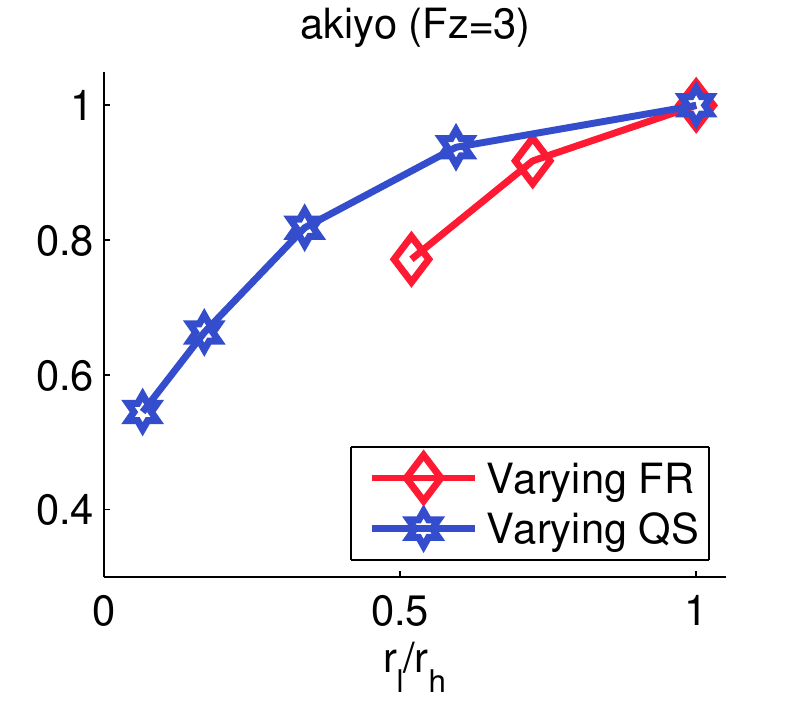}
  \includegraphics[scale=0.41]{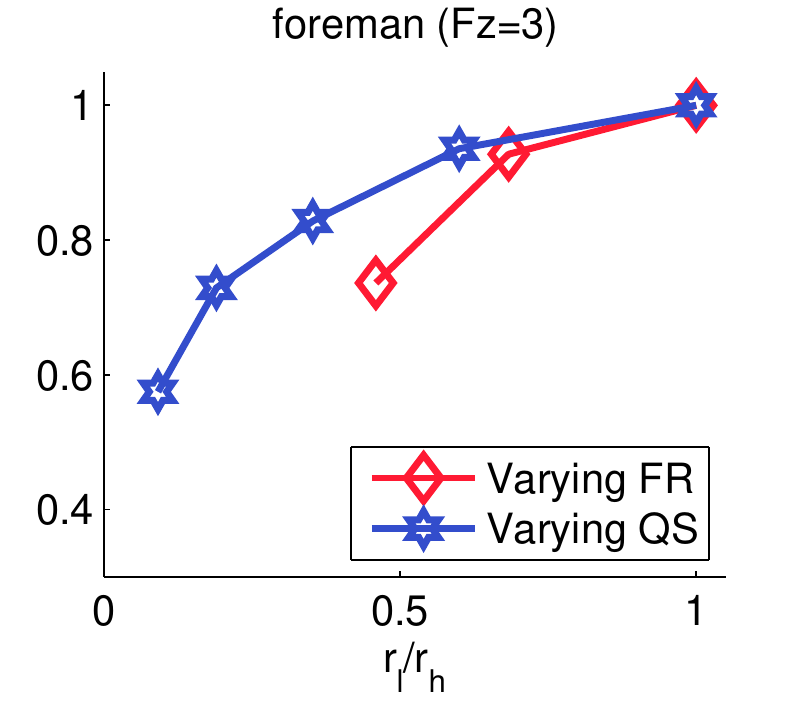}
  \includegraphics[scale=0.41]{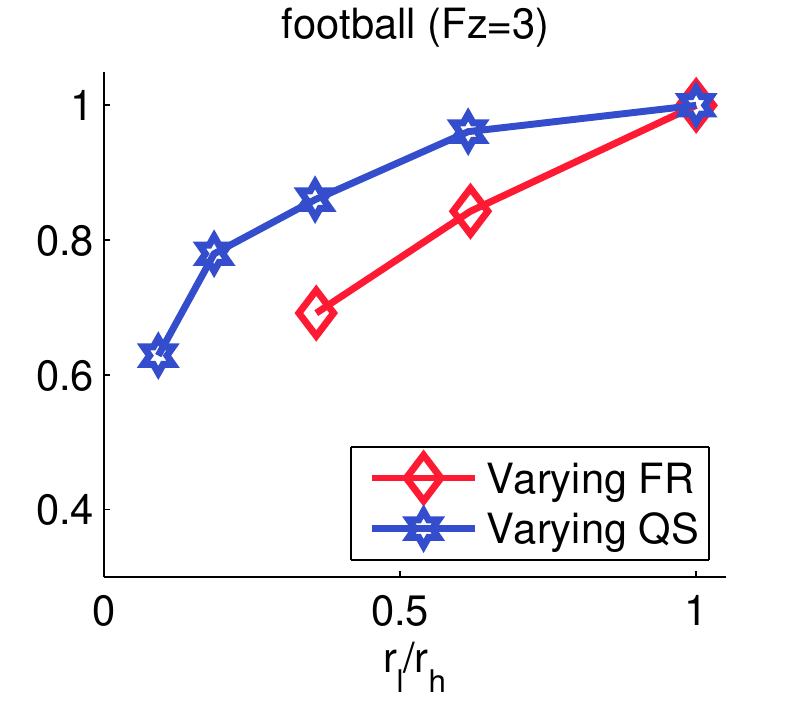}
  \includegraphics[scale=0.41]{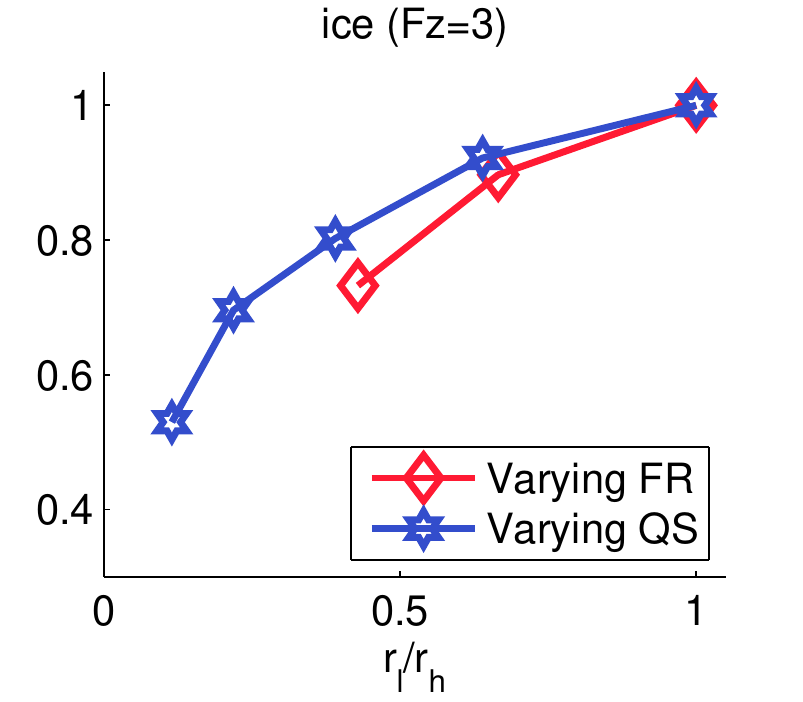}
  \includegraphics[scale=0.41]{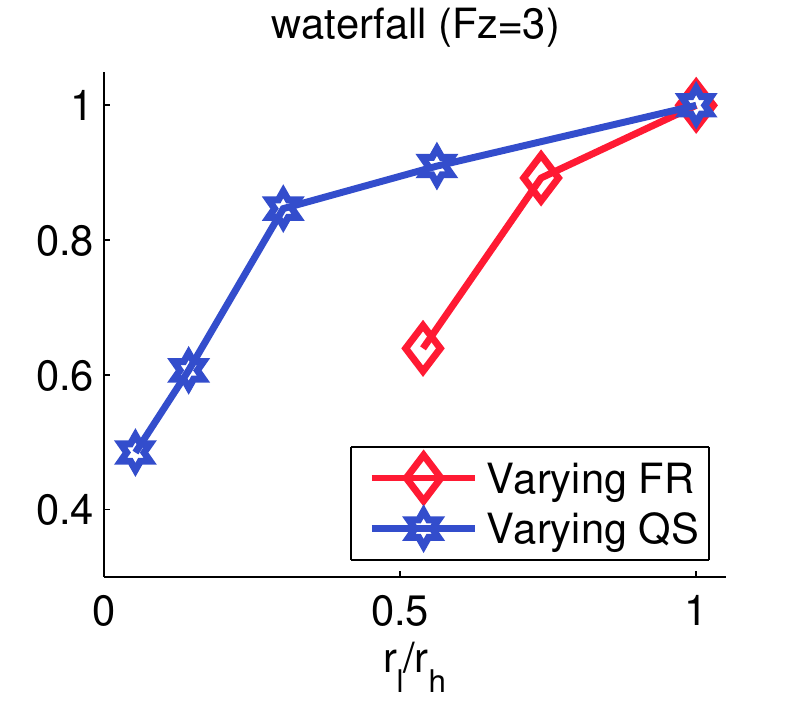}
 \caption{Q($r_{h},r_{l}$) vs. $r_{l}/r_{h}$ when videos are coded using H.264/SVC. Blue points are for videos obtained by varying FR, whereas red points for obtained by varying FR. The high rate point is obtained with QS $q_l$=$16$, $t_h$=$30$. Points are the measured data with different markers and colors corresponding to different $r_h$ at $q=16$ or $t=30$Hz.}
 \label{fig:FRQPVar_cmp_SVC}
\end{figure*}


\section{Adaptation of FR and QS under the Same Rate Variation}\label{sec:FRQPVar_cmp}
Recall that the motivation for this study is how to adapt a video when the available bandwidth changes. A natural question to ask is, given the rate variation pattern,  which encoding parameter should we change, QS or FR? To answer this question, we find the bit rates corresponding to different QS and FR variations, and replot the quality in terms of the same bit rate variation.


Let $r_{l}$ and $r_{h}$ denote the bitrate at $t_l$ (or $q_h$) and $t_h$ (or $q_l$), respectively. We plot the measured data, defined as Q($r_{h},r_{l}$) = MOS($r_{h},r_{l}$)/MOS($r_{\max},r_{\max}$) in terms of bitrate ratio $r_{l}/r_{h}$ in Figs.~\ref{fig:FRQPVar_cmp_SVC}  , where the high rate point ($r_h$) is achieved with FR=30 and QS=16. The rates corresponding to different FR and QS combinations are the actual bit rates when the test videos are coded using the H.264/SVC encoder using the configuration described in Sec.~\ref{sec:SQmeasurement}. Because the quality ratings between different Fz's are quite similar, so we only present the cases with Fz=3 only. Figure.~\ref{fig:FRQPVar_cmp_SVC} demonstrates the quality degradation v.s. rate ratio. It is clear that for the same periodic rate variation, videos with QS variation are more preferable than FR variation.
We would like to stress that this result is for a particular  high rate point when the video is coded at a very high quality. When the video is already at a relatively low rate (coded using a lower FR or higher QS), it is possible that lowering FR may be better than increasing QS, if one has to further reduce the bit rate periodically. Unfortunately, we don't have subjective ratings for videos starting at a common encoding configuration at a low bit rate. Acquiring such data is beyond the scope of this paper.

\section{Statistic Analysis}\label{sec:ANOVA_test}
The results presented in Sec.~\ref{sec:Test_results_FRVar} and~\ref{sec:Test_results_QPVar} show that FR/QS variation magnitude, variation frequency and video content all affect the perceptual quality. To evaluate whether the changes in quality ratings due to these factors and their interactions are statistically significant, we perform three-way Analysis of Variance (ANOVA)~\cite{Snedecor_Statistics}. With ANOVA, we compute the probability ($p$-value, which is derived from the cumulative distribution function of F based on the $F$-value) of the event that the difference in MOS when a particular variable is changed is due to chance. If this probability is low ($p$-value $<$ 0.05), we consider this variable as having statistical significance on MOS.

\begin{table}[!htp]
\centering
\caption{Three-way ANOVA results for FR variation using Q($t_h$,$t_l$) data}\label{tab:ANOVA_FRVar}
\begin{tabular}{@{~}c@{ } p{1.5 cm}|c|c|c}
\hline
& Cases & Factors & $F$-value & $p$-value\\
\hline
\hline
\multirow{3}{*}{${\rm P_1}$:} & \multirow{3}{*}{All Data}&$\Delta$t &  26.3  & 0 \\
&&Content &  83.83  & 0  \\
&&Frequency &   20.62 & 0  \\
\hline
\hline
${\rm P_2}$: & (30, 15) and (30, 30) &$\Delta t$ & 1.74  & 0.187 \\
\hline
\hline
${\rm P_3}$: & (15, 7.5) and (15, 15) &$\Delta t$ & 45.02  & 0 \\
\hline
\hline
${\rm P_4}$: & (30, 15) and (15, 15) &$\Delta t$ & 44.52  & 0 \\
\hline
\hline
${\rm P_5}$: & (15, 7.5) and (7.5, 7.5) &$\Delta t$ & 31.36  & 0 \\
\hline
\hline
\end{tabular}
\end{table}

\begin{table}[!htp]
\centering
\caption{The ANOVA test for the FR variation frequency using Q($t_h,t_l$)/Q($t_h,t_h$) data}\label{tab:FRVar_ANOVA_Fz}
\begin{tabular}{c|c|c|c}
\hline
Target pair & Factors & $F$-value & $p$-value\\
\hline
\hline
\multicolumn{4}{c}{all Fz's and $t_h$'s}\\
\hline
 & $\Delta t$*Fz &  0.88 & 0.54  \\
\hline
\multicolumn{4}{c}{when $t_h$ = 30Hz, $t_l$=30, 15, 7.5}\\
\hline
Fz = 1 and 2 & $\Delta t$*Fz & 1.76 & 0.19 \\
Fz = 1 and 3 & $\Delta t$*Fz & 1.44  & 0.25  \\
Fz = 2 and 3 & $\Delta t$*Fz &  0.49 &  0.61 \\
Fz =1, 2, 3 & $\Delta t$*Fz & 1.28  & 0.29  \\
\hline
\hline
\multicolumn{4}{c}{when $t_h$ = 15Hz, , $t_l$=15, 7.5}\\
\hline
Fz = 1 and 2 & $\Delta t$*Fz &  3.43 & 0.08  \\
Fz = 1 and 3 & $\Delta t$*Fz & 3.41  & 0.08  \\
Fz = 2 and 3 & $\Delta t$*Fz & 0.001  & 0.96  \\
Fz =1, 2, 3 & $\Delta t$*Fz & 2.44 & 0.10  \\
\hline
\end{tabular}
\end{table}

\subsection{Analysis for Frame Rate Variation}\label{ssec:FRVar_ANOVA_test}
In Tab.~\ref{tab:ANOVA_FRVar}, results given under ${\rm P_1}$ are obtained by considering all test conditions together. We can see that FR variation magnitude (indicated by $\Delta t$=$t_h-t_l$), variation frequency (inversely related to Fz), and video content all have statistically significant impact on the subjective ratings. 
In addition to conducting ANOVA over all data, we also looked at the significance of variation magnitude for a few specific cases. The impact of FR variation is significant in all the cases except ${\rm P_2}$. This means that the quality rating difference between a video with constant FR of 30 Hz, and that alternating between 30 and 15Hz, is mostly due to viewers' inconsistency.



Although Tab.~\ref{tab:ANOVA_FRVar} shows that Fz has a significant impact on the absolute quality rating Q($t_h,t_l$),
we are interested to know the statistical significance of the influence of Fz on the dropping trend of Q($t_h,t_l$)/Q($t_h,t_h$) data, as mentioned in Sec.~\ref{ssec:Imp_FRVar}. Toward this goal, we conduct a two-way repetition ANOVA test on the Q($t_h,t_l$)/Q($t_h,t_h$) data. Results in Tab.~\ref{tab:FRVar_ANOVA_Fz} suggest that there is no significant interaction between FR variation magnitude and Fz on the Q($t_h,t_l$)/Q($t_h,t_h$) data, with $p$-value $>$ 0.05 under all combinations of $t_h$ and $t_l$ examined.

\subsection{Statistical Significance for QS Variation}\label{ssec:QPVar_ANOVA_test}
Results given under ${\rm P_1}$ in Tab.~\ref{tab:ANOVA_QPVar} are obtained by considering all test conditions together. We can see that QS variation magnitude (indicated by $\Delta q$=$q_h-q_l$), variation frequency, and video content all have significant impact on the subjective ratings.
In addition to conducting ANOVA over all data, we also looked at a few specific cases, each comparing ratings for two ($q_l$, $q_h$) pairs. The ANOVA results show that, when $q_l$ and $q_h$ are similar, e.g., ${\rm P_2}$ and ${\rm P_3}$, the quality difference due to QS variation is insignificant, which is as expected. In particular, this result confirms that the observed change when switching from $q_h$=40 to $q_l$=25 or 16 in Fig.~\ref{fig:NMOS_ql_qh_QP_h} is not statistically significant.

To further investigate the dependency of the dropping trend of Q($q_l,q_h$)/Q($q_l,q_l$) on Fz, we conducted a two-way ANOVA test on Q($q_l,q_h$)/Q($q_l,q_l$) data. From Tab.~\ref{tab:QPVar_ANOVA_Fz}, we learn that there is no significant difference between Fz=1 and 2 nor between Fz=2 and 3, but the difference between Fz=1 and Fz=3 is significant. This results is quite consistent with the observations from Fig.~\ref{fig:QPVar_NMOS_ratioQS_invexp_4par}, and this is the basis we choose to use the same model parameter $\alpha_{qv}$ for Fz=1 and Fz=2, but a different value for Fz=3 in the ${\rm QQV}$ model in~(\ref{eq:QPVar_overallQ}).
\begin{table}[!ht]
\centering
\caption{Three-way ANOVA results for QS Variation using Q($q_l,q_h$) data}\label{tab:ANOVA_QPVar}
\begin{tabular}{@{~}c@{ } p{1.9 cm}|c|c|c}
\hline
& Cases & Factors & $F$-value & $p$-value\\
\hline
\hline
\multirow{3}{*}{${\rm P_1}$:} & \multirow{3}{*}{All Data}&$\Delta q$ &  26.8 & 1e-5  \\
 &&Content &  5.25  & 9e-4  \\
 &&Frequency &  4.28  & 0.01  \\
\hline
\hline
${\rm P_2}$:& Q(40, 40) and Q(40, 25)&$\Delta q$ & 0.02  & 0.886  \\
\hline
\hline
${\rm P_3}$:& Q(40, 40) and Q(40, 16)&$\Delta q$ & 2.53  & 0.15  \\
\hline
\hline
${\rm P_4}$:& Q(16, 16) and Q(16, 102)&$\Delta q$ & 3041  & 0  \\
\hline
\hline
${\rm P_5}$:& Q(102, 102) and Q(16, 102)&$\Delta q$ & 368.79  & 0  \\
\hline
\hline
\end{tabular}
\end{table}

\begin{table}[!htp]
\centering
\caption{The ANOVA test for the QS variation frequency using Q($q_h,q_l$)/Q($q_l,q_l$) data}\label{tab:QPVar_ANOVA_Fz}
\begin{tabular}{c|c|c|c}
\hline
Target pair & Factors & $F$-value & $p$-value\\
\hline
\hline
\multicolumn{4}{c}{all Fz's and $q_l$'s}\\
\hline
 & $\Delta q$*Fz &  1.22 & 0.27  \\
\hline
\multicolumn{4}{c}{when $q_l$ = 28}\\
\hline
Fz = 1 and 2 & $\Delta q$*Fz &  0.56 & 0.69  \\
Fz = 1 and 3 & $\Delta q$*Fz &  2.37 & 0.06  \\
Fz =  2 and 3 & $\Delta q$*Fz &  1.38 & 0.25  \\
Fz =1, 2, 3 & $\Delta q$*Fz & 1.61 & 0.19  \\

\hline
\hline
\multicolumn{4}{c}{when $q_l$ = 36}\\
\hline
Fz = 1 and 2 & $\Delta q$*Fz &  0.19 & 0.82  \\
Fz = 1 and 3 & $\Delta q$*Fz &  3.85 & 0.03  \\
Fz =  2 and 3 & $\Delta q$*Fz &  1.54 & 0.23  \\
Fz =1, 2, 3 & $\Delta q$*Fz & 1.61 & 0.19  \\
\hline
\end{tabular}
\end{table}



%
%

\section{Conclusion}\label{sec:Conclusion}
In this paper, we report the results of our subjective experiments to investigate the impact
of periodic FR/QS variation on the perceived video quality. We observed following interesting trends.
Regarding the FR variation, firstly, under the same average FR the quality for a video with a constant frame rate is higher than that with FR variation, alternating between $t_l$ and $t_h$; and Secondly, the degradation due to FR change is more severe when $t_h/t_l$ ratio is higher, especially when $t_h>$ 2$t_l$;
Thirdly, alternating between $t_l$ and $t_h$ is generally better than staying at $t_l$. However, the quality improvement become saturated when $t_h/t_l > 2$. Finally the variation frequency does not have significant impact on the quality decay relative to the quality with a constant FR equal to $t_h$. The last two observations are   somewhat surprising, and may not be true when Fz $<$ 1 sec.

Similar phenomena have also been observed under QS variation. Firstly, under the same average QS, a video with a constant QS is perceptually more appealing than a video with variable QS under the same average QS. Secondly, the degradation due to QS change is more severe when constant low QS $q_l$ is relatively low and $q_h/q_l$ ratio is high. Thirdly, alternating between $q_h$ and $q_l$ is generally equal or better than a video with a constant high QS $q_h$. However, the improvement becomes saturated when $q_l/q_h<$ 0.4.
Unlike the case with FR variation, the variation frequency does affect the quality decay relative to the quality with a constant QS equal to $q_l$. Overall, slower variation leads to less quality decay, as expected.

These results provide several important guidelines for rate adaptation in video transmission over dynamically changing networks. For example, when the underlying application allows sufficiently long delay, it is better to code a video at constant FR and QS, and use a large buffer at the sender to deal with the bandwidth variation. When the delay constraint is stringent and one cannot afford to use large buffers, at least in the range of variation frequency considered here (Fz $>=$ 1 sec), it is better to allow the rate to vary with the available bandwidth (by adjusting FR or QS), rather than staying at the lower rate. This suggests that trying to keep a smooth but low video quality may not be the right strategy for video streaming in dynamic networks. Allowing  a certain amount of quality fluctuation may in fact lead to better quality.
However, the system should limit the amount of changes in FR or QS. Specifically, when the lowest bandwidth demands a FR=$t_l$ and QS=$q_h$, the system should limit the FR to at most 2$t_l$, and QS to at least 0.4$q_h$, even if the instantaneous bandwidth allows higher $t$ or lower $q$ occasionally. When the rate reduction can be made by changing either QS or FR, it is suggested that increasing QS is preferred than decreasing FR to achieve the least quality degradation. By comparing video quality under the same periodic rate variation, we further found that adapting QS is better than changing FR, when the high rate point allows the video to be coded at very good quality (e.g. with high FR and low QS). Somewhat surprising, this is observed for video with either low or high motion.


Based on the observed relation between the quality and FR or QS variations, we propose several quality models. For FR variation, the model ${\rm QTV}$ determines the quality of a video with alternating frames rates of $t_h$ and $t_l$ as the product of two terms. The first term reflects the quality of a video at a constant FR of $t_h$ and the second term indicates the degradation due to the switching between $t_h$ and $t_l$, and depends only on the ratio $t_l/t_h$. Similarly, we have developed ${\rm QQV}$ model for videos with QS variation. We note that these models are proposed based on the trends observed from the test sequences used in our subjective tests, and they need to be validated using other test sequences. In this study, the model parameters are derived using least square fitting of the model with the mean opinion scores for each sequence. For the model to be useful, one must also study how to predict the model parameters from the video content. Furthermore, rather than predicting the model parameter $\alpha_t$ for each $t_h$, it may be possible to find a simple functional form to model  $\alpha_t$($t_h$). Similarly, one may examine and model the underlying relation of $\alpha_q$ with $q_h$ and Fz. These are interesting directions for future studies.



This paper focuses only on the impact of periodic frame rate/QS changes on the perceptual quality. One challenging issue in modeling the impact of FR and QS variation on the quality is how to design the subjective test to help understand the effect of infinitely many likely variation patterns. Our subjective study that considers only periodic variations is only the first step towards this challenge, and is motivated by the fact that any pattern can be decomposed into periodic patterns with different frequencies. Our study so far considered only the effect of FR and QS variation, individually, and under a few variation frequencies. Future studies may examine a wider range of variation frequency, variation of spatial resolution, and joint impact of FR, QS and spatial resolution variation. 


\bibliographystyle{IEEEtran}

\end{document}